\documentclass[journal]{IEEEtran}
\IEEEoverridecommandlockouts
\usepackage{lineno}
\usepackage{hyperref}
\usepackage{cite}
\usepackage{amsmath,amssymb,amsfonts,cases} 
\usepackage[linesnumbered,ruled,vlined]{algorithm2e}
\usepackage{amsmath}
\usepackage{amsthm}

\usepackage{graphicx}
\usepackage{textcomp}
\usepackage{xcolor}
\usepackage{graphicx}
\usepackage{amsmath,mleftright}
\usepackage{amsfonts,amssymb}
\usepackage{mathrsfs}
\usepackage{mathtools}
\usepackage{bm}
\usepackage{multirow}
\usepackage{array}
\usepackage{amssymb}
\usepackage{amsmath}
\usepackage{cite}
\usepackage{url}
\usepackage{xcolor}
\usepackage{booktabs}
\usepackage{tabularx}
\usepackage{cite,graphicx,amsmath,amssymb}
\usepackage{subfig}
\usepackage{graphicx}
\usepackage{fancyhdr}
\usepackage{mdwmath}
\usepackage{mdwtab}
\usepackage{caption}
\usepackage{amsthm}
\usepackage{setspace}
\usepackage{bm}
\usepackage{mathtools}
\usepackage{dsfont}
\usepackage{bbm}
\usepackage{framed}

\newtheorem{theorem}{Theorem}

\newtheorem{lemma}{Lemma}

\newtheorem{corollary}{Corollary}

\makeatletter
\newcommand{\biggg}{\bBigg@{3}}
\newcommand{\Biggg}{\bBigg@{3.5}}
\makeatother
\makeatletter
\renewcommand{\maketag@@@}[1]{\hbox{\m@th\normalsize\normalfont#1}}%
\makeatother
\def\BibTeX{{\rm B\kern-.05em{\sc i\kern-.025em b}\kern-.08em
		T\kern-.1667em\lower.7ex\hbox{E}\kern-.125emX}}
\expandafter\def\expandafter\normalsize\expandafter{%
	\normalsize%
	\setlength\abovedisplayskip{4pt}%
	\setlength\belowdisplayskip{4pt}%
	\setlength\abovedisplayshortskip{2pt}%
	\setlength\belowdisplayshortskip{2pt}%
}

\begin{document}
\title{Pinching Antenna System (PASS) Enhanced Covert Communications: Against Warden via Sensing}

\author{Hao Jiang, \IEEEmembership{Student Member,~IEEE}, Zhaolin Wang, \IEEEmembership{Member,~IEEE}, Yuanwei Liu, \IEEEmembership{Fellow,~IEEE}, \\
Arumugam Nallanathan, \IEEEmembership{Fellow,~IEEE}, and Zhiguo Ding, \IEEEmembership{Fellow,~IEEE}
\thanks{H. Jiang, and A. Nallanathan are with the School of Electronic Engineering and Computer Science, Queen Mary University of London, London E1 4NS, U.K. (e-mail: \{hao.jiang, a.nallanathan\}@qmul.ac.uk).}
\thanks{Z. Wang and Y. Liu are with the Department of Electrical and Electronic Engineering, The University of Hong Kong, Hong Kong (e-mail: \{zhaolin.wang, yuanwei\}@hku.hk).}
\thanks{Z. Ding is with Khalifa University, Abu Dhabi, UAE, and the University of Manchester, Manchester, M1 9BB, U.K. (e-mail: zhiguo.ding@manchester.ac.uk).}
\vspace{-10pt}
}

\maketitle

\begin{abstract}
A sensing-aided covert communication network empowered by pinching antenna systems (PASS) is proposed in this work.
Unlike conventional fixed-position MIMO arrays, PASS dynamically reconfigures its pinching antennas (PAs) closer to the legitimate user, substantially enhancing covertness. 
To further secure the adversary’s channel state information (CSI), a sensing function is leveraged to track the malicious warden’s movements.
In particular, this paper first proposes an extended Kalman filter (EKF)–based approach to fulfilling the tracking function.
Building on this, a covert communication problem is formulated with a joint design of beamforming, artificial noise (AN) signals, and the position of PAs.
Then, the beamforming and AN design subproblems are resolved jointly with a subspace approach, while the PA position optimization subproblem is handled by a deep reinforcement learning (DRL) approach by treating the evolution of the warden's mobility status as a temporally corrected process.
Numerical results are presented and demonstrate that: i) the EKF approach can accurately track the warden's CSI with low complexity, ii) the effectiveness of the proposed solution is verified by its outperformance over the greedy and searching-based benchmarks, and iii) with new design degrees of freedom (DoFs), the performance of PASS is superior to the conventional fully-digital MIMO systems.
\end{abstract}

\section{Introduction}
With the commercialization of the fifth generation (5G) wireless network on a global scale \cite{dang2020what, you2021towards}, the next generation wireless communication technology is around the corner. 
It is anticipated to offer revolutionary improvements on the key performance indicators (KPIs), such as data rate, coverage, and energy efficiency, etc \cite{jiang2021the}.
In the 5G era, both the performance of modern communication technologies and the volume of signals they deliver have met user demands, suggesting that further efforts focused solely on increasing throughput may yield diminishing returns.
Therefore, next-generation communication technologies are expected to explore a broader spectrum of wireless-signal functionalities.

In recent years, sensing functionalities of wireless signals have drawn significant attention from both academia and industry, heralding a future of integrated sensing and communications (ISAC) networks \cite{liu2022integrated, cong2024near}.
The ISAC framework enables the shared use of resource blocks between the sensing and communication functionalities, thereby reducing energy consumption and hardware costs.
ISAC operation generally falls into two paradigms: i) \emph{Communication-Aided Sensing}, where the reflected communication waveform is re-exploited to probe the sensing target, and ii) \emph{Sensing-Aided Communication}, where the environment awareness attained from sensing is fed back to strengthen the communication function.
By incorporating both functionalities with distinct purposes into a single synergetic system, the first paradigm introduces the communication-sensing tradeoff \cite{liu2022survey}, implying that enhancing one functionality will inevitably sacrifice the other.
This communication-sensing tradeoff is undesirable in communication-centric networks.
Hence, real-world deployments increasingly gravitate toward the second paradigm, which emphasizes communication performance by placing sensing in a serving role.
This paradigm has been exploited for beam training \cite{jiang2024sense}, beam tracking \cite{liu2020radar}, and channel estimation \cite{liu2025sensing}, etc.
In fact, the basic idea of sensing-aided communications is to use sensing as a perception tool to obtain information about the unknown environment. 
In physical layer security (PLS) systems, the channel state information (CSI) of the adversary is typically difficult to obtain at the transmitter \cite{he2014mimo}, due to the non-cooperation of the malicious user during the channel estimation (CE) stage.
However, the physical parameters that determine the adversary’s CSI can be inferred from the signals it reflects passively.
This fact motivates the use of sensing to assist PLS \cite{wei2022toward}.

Compared with sensing-assisted pure communications, research on the sensing-assisted PLS remains in its infancy.
The very first work on this topic was \cite{deligiann2018secrecy}, which highlighted security issues in the ISAC systems by treating the communication user as the legitimate receiver and the radar target as the malicious eavesdropper. 
In that work, the authors maximized the secrecy rate while ensuring the target-detection threshold was met.
However, here comes a design dilemma:
To meet the sensing threshold, one can allocate more power toward the eavesdropper’s location; however, this inevitably increases the risk of confidential information leakage.
Fortunately, artificial noise (AN) can be explored as a remedy to address this dilemma.
In pure PLS systems, where sensing is not considered, AN is solely used to jam malicious users, thereby worsening their signal-to-interference-plus-noise ratio (SINR).
In ISAC PLS systems, AN serves a dual role: it continues to jam eavesdroppers while also acting as a probing signal for the sensing of eavesdroppers. 
By integrating AN for both jamming and sensing purposes, the sensing requirements can be satisfied without compromising confidentiality.
Building on this idea, the authors in \cite{su2021secure} considered a secure sensing-communication system under the perfect and imperfect CSI at the eavesdropper, where the beam pattern optimization was considered.
As a separate research, the authors in \cite{jia2024physical} considered a similar system setup, using the Cramér–Rao bound (CRB) as the sensing performance metric, which was more closely related to detection accuracy.
In both works, however, knowledge of the adversary’s CSI remains indispensable, thereby offering valuable performance benchmarks yet posing practical challenges for real-world deployments.
To alleviate the dependency on the adversary's CSI, the following research has been conducted by integrating the detection procedure.
Considering a similar setup with \cite{su2021secure}, the authors in \cite{su2024sensing} introduced a two-stage procedure: in the first stage, the omni-directional beam was emitted to localize the eavesdroppers, while in the second stage, the CRB was optimized based on the detection results and the PLS design was conducted using the sensing results, further improving the sensing accuracy.
Their results also showcased that the directional design of AN can offer a high secrecy rate gain compared to isotropic AN designs without the aid of sensing.

To further enhance the security of ISAC, covert communications have emerged as a new type of PLS techniques \cite{chen2023covert}.
In contrast to conventional PLS, which minimizes information leakage to malicious users, covert communication aims to hide the legitimate transmission from detection by the eavesdroppers.
Therefore, the confidentiality level of communication systems will not be compromised by eavesdroppers with improved ability to decode \cite{jin2025covert, bash2015hiding}.
To fulfill this design goal, covert communication designs are targeted at increasing the total error rate for the adversary, which includes both the false-alarm and mis-detection probabilities.
By leveraging on AN to falsely ``alarm" malicious detection, the total error rate will increase, thus enhancing the covertness of the legitimate transmission \cite{soltani2018covert}.
Thus, similar to exploring AN for both jamming and sensing purposes in conventional PLS systems, sensing-aided covert communication with AN also attracts substantial attention from the community.
Based on pioneering work \cite{wei2022safeguard} that considered a conventional PLS setup, the authors in \cite{wang2024sensing} utilized the sensing signals to interfere with the malicious user while tracking its trajectory, exploring the possibility of ``turning interference into allies."
Building on this work, a two-stage covert communication framework was proposed by \cite{qian2025two}.
As a separate study, \cite{tang2025dual} utilized a dual-functional AN (DFAN) to combat an adversary warden in the ISAC system.

Although the advantages of sensing in PLS systems using MIMO have been strongly demonstrated by the above research endeavors, the higher requirements of next-generation communication networks have motivated us to explore new types of reconfigurable antennas.
Compared to conventional MIMO, reconfigurable antennas manipulate the wireless channel in a desirable way, thereby enhancing channel capacity \cite{ding2025flexible}.
Existing antenna types, such as fluid antennas \cite{new2024tutorial} and movable antennas \cite{zhu2025tutorial}, mainly offer wavelength-scale repositioning to reconfigure small-scale fading, which limits the reconfiguration capabilities of these antennas.
To overcome this limitation, NTT DOCOMO Inc. proposed the Pinching Antenna System (PASS), with the prototype demonstrated in 2021 \cite{suzuki2022pinching}. 
From a manufacturing standpoint, PASS consists of multiple low‐attenuation waveguides equipped with dielectric pinching antennas (PAs) that emit signals into free space.
Since these waveguides can extend over several meters, PASS enables reconfiguration of both small‐scale and large‐scale fading, and even establishes line‐of‐sight propagation conditions \cite{liu2025pinching}, thus sharing a similar concept with surface wave communication (SWC) \cite{wong2021vision}. 
This capability has proven advantageous for security-focused communications, enhancing both conventional PLS systems \cite{sun2025physical} and covert communication networks \cite{jiang2025pinching}.

However, these works all assume that perfect or imperfect CSI is accessible at the transmitter, thus diminishing the practicality of these works.
They also treat adversaries as stationary, which may not reflect real‐world scenarios. 
Inspired by recent progress in sensing-aided PLS for conventional antennas, we leverage PASS’s integrated sensing capability to estimate the CSI of a moving malicious eavesdropper in a covert communication setting
This is a more realistic yet challenging scenario highlighted in \cite{chen2023covert}.
The main contributions of this paper are summarized in what follows:
\begin{itemize}
    \item We propose a sensing-aided covert communication system empowered by PASS.
    In particular, the transmitter (Alice) intends to confidentially transmit information to the legitimate receiver (Bob) while evading detection by a malicious mobility warden (Willie).
    To acquire the CSI of Willie, the sensing functionality is leveraged to extract the real-time mobility status of Willie from echo signals.
    Additionally, the AN is utilized both to jam Willie and to enhance sensing performance.
    
    \item We develop an extended Kalman filter (EKF) method for tracking the full mobility status of Willie, including distance, angle, amplitude velocity, and direction velocity. 
    This full-dimensional sensing is enabled by near-field effects caused by the large aperture of PASS.

    \item We formulate a covert communication optimization problem to maximize the covert rate at Bob by tuning the PA's position, beamforming, and AN signal, subject to Willie's total error constraint and the sensing accuracy requirements.
    Leveraging on the sensing results, the unknown parameters regarding Willie's CSI can be obtained.
    To enable real-time optimization in this dynamic system, we first devise a low-complexity optimal solution to the beamforming and AN design subproblem.
    Then, a deep reinforcement learning (DRL) approach is utilized to optimize PA's positions, exploiting the temporal correlations in Willie's movement.
    
    \item We provide numerical results to validate the advantages of PASS and the effectiveness of the proposed algorithm. 
    The results demonstrate that i)~PASS can achieve high-fidelity sensing via a small number of RF chains; 
    ii)~ The proposed approach gradually learns to adjust the position of PAs with a much lower complexity than an exhaustive search and a heuristic greedy solution;
    ii)~PASS outperforms the conventional fixed-position MIMO systems, although the perfect mobility status of Willie is accessible for the MIMO benchmark.
\end{itemize}

The remainder of the paper is organized as follows. 
Section \ref{sect:system_model} presents the ISAC–PLS system model. 
Section \ref{sect:malicious_tracking} details the sensing-assisted framework for covertness design, followed by a problem formulation. 
Section \ref{sect:solution} elaborates on the joint beamforming, AN, and PA‐positioning solution to the formulated problem. 
Numerical results are provided in Section \ref{sect:results}, and conclusions are drawn in Section \ref{sect:conclusion}.

\textit{Notations:}
Scalars, vectors, and matrices are denoted by the lower-case, bold-face lower-case, and bold-face upper-case letters, respectively.
$\mathbb{C}^{M \times N}$ and $\mathbb{R}^{M \times N}$ denote the space of $M \times N$ complex and real matrices, respectively.
$(\cdot)^\mathrm{T}$, $(\cdot)^*$, and $(\cdot)^\mathrm{H}$ denote the transpose, conjugate, and conjugate transpose, respectively.
$|\cdot|$ represents absolute value.
For a vector $\mathbf{a}$, $[\mathbf{a}]_i$, $\left\| \mathbf{a} \right\|_1 $, and $\left\| \mathbf{a} \right\|_2 $ denote the $i$-th element, $1$-norm, and $2$-norm, respectively.
$\odot$ and $\left<\cdot,\cdot \right>$ denote the element-wise product and the inner product, respectively. 
$\mathrm{j}=\sqrt{-1}$ denotes the imaginary unit.

\section{System Model} \label{sect:system_model}
\begin{figure}[t!]
    \centering
    \includegraphics[width=0.8\linewidth]{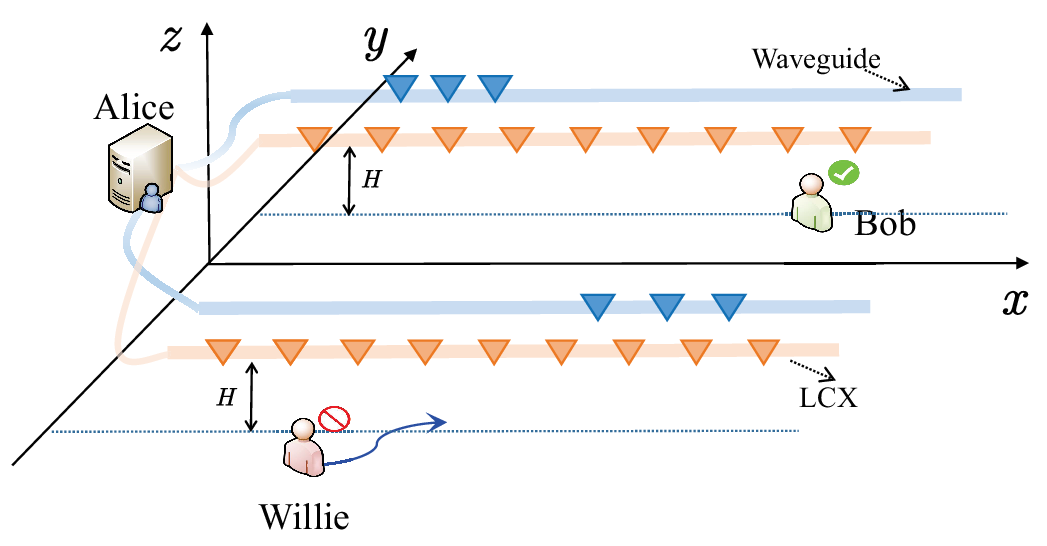}
    \caption{Illustration of a sensing-aided covert-communication system using PASS, featuring a mobile adversary Willie.}
    \label{fig:system_model}
    \vspace{-10pt}
\end{figure}
We consider a sensing-aided covert-communication system using PASS, which is illustrated in Fig. \ref{fig:system_model}.
According to the classic covert communication setup \cite{chen2023covert}, the base station (BS), referred to as Alice, aims to convey confidential information to the single-antenna legitimate user, referred to as Bob, without being detected by the single-antenna adversary warden, referred to as Willie. 
However, due to its malicious nature, Willie will not share the CSI with Alice, making it challenging to guarantee the transition's covertness.
Moreover, in this work, we consider a more tricky scenario, where Willie moves on the $xy$ plane while Bob is stationary. 
To address this issue, the ISAC technology is utilized at Alice, which transmits confidential information to Bob while acquiring the CSI of Willie via sensing.

Accordingly, for the communication functionality, Alice is equipped with a PASS featuring $N_{\rm t}$ waveguides, each of which is fed by a dedicated RF chain. 
The signal in each waveguide is then radiated to free space by $M_{\rm t}$ attached PAs.
The $x$-coordinate of the fed points of PASS is set to $x=0$.
For the sensing functionality, $N_{\rm r}$ leaky coaxial cables (LCX) are utilized for echo reception \cite{ji2024theoretical}, each with $M_{\rm r}$ uniformly spaced slots for echo signal reception.
Both the transmit waveguides and receive LCXs are placed in parallel to the $x$-axis at a height of $H$ with a uniform spacing of $D$.
In addition to the above, the lengths of waveguides and LCXs are uniformly set to $L_{\max}$. 
Time is slotted into coherent time intervals (CPIs) indexed by $t$, during which Willie's mobility status can be regarded as unchanged.
Below, we present the communication signal model, the sensing signal model, and the covertness model. 
For notational simplicity, we omit the CPI index $t$ in this section.

\subsection{Communication Signal Model} \label{sect:communication model}
Let the position of the $m$-th PA on the $n$-th waveguide and the position of Bob be denoted by $\mathbf{p}_{\mathrm{t}, n,m}=[x_{\mathrm{t}, n, m}, y_{\mathrm{t}, n, m}, H]^{\mathrm{T}}\in \mathbb{R}^{3 \times 1}$ and $\mathbf{r}_{\rm{b}} = [x_{\mathrm{b}}, y_{\mathrm{b}}, 0]^{\mathrm{T}} \in \mathbb{R}^{3 \times 1} $, respectively.
Thus, the distance between the $m$-th PA on the $n$-th waveguide and Bob can be computed via
\begin{align}
    r_{\mathrm{b}, n, m} = \| \mathbf{p}_{\mathrm{t}, n,m} - \mathbf{r}_{\rm{b}} \|_2.
\end{align}
Letting $c_n\in\mathbb{C}^{1 \times 1}$ be the $n$-th entry of the signal $\mathbf{c} \in \mathbb{C}^{N_{\rm t} \times 1}$, the overall received signal at Bob is given by
\begin{align}
    y_{\mathrm{b}}=\sum_{n=1}^{N_{\mathrm{t}}}{\mathbf{a}_{\mathrm{b},n}^{\mathrm{H}}\left( \mathbf{r}_{\mathrm{b}},\mathbf{x}_n \right) \mathbf{g}_n\left( \mathbf{x}_n \right) c_n+n_{\mathrm{b}}}, \label{eq:bob_channel}
\end{align}
where $n_{\rm b} \sim \mathcal{CN}(0, \sigma^2_{\rm b})$ denotes the additive complex-valued Gaussian noise with zero mean and power of $\sigma^2_{\rm b}$, $\mathbf{x}_n  \triangleq [x_{n, 1}, ..., x_{n, M_{\mathrm{t}}}]^{\mathrm{T}} \in \mathbb{R}^{M_{\rm t} \times 1}$ denotes the $x$-coordinate vector of the PAs on the $n$-th waveguide, $\mathbf{g}_n\left( \mathbf{x}_n \right) \in \mathbb{C}^{ M_{\mathrm{t}} \times 1}$ denotes the in-waveguide channel vector, and $\mathbf{a}_{\mathrm{b}, n}\left(\mathbf{r}_{\rm b}, \mathbf{x}_n  \right) \in \mathbb{C}^{ M_{\mathrm{t}} \times 1}$ denotes the free-space channel vector.
In particular, arising from the propagation between the feed point and PAs, the in-waveguide channel vector can be expressed 
\begin{align}
    \mathbf{g}_n\left( \mathbf{x}_n \right) =\left[ \alpha _1e^{-\mathrm{j}k_{\mathrm{g}}x_{\mathrm{t},n,1}},...,\alpha _{M_{\mathrm{t}}}e_{}^{-\mathrm{j}k_{\mathrm{g}}x_{\mathrm{t},n,M_{\mathrm{t}}}} \right] ^{\mathrm{T}}, \label{eq:in-waveguide}
\end{align}
where $k_{\mathrm{g}} \triangleq 2 \pi / \lambda_{\mathrm{g}}= 2 \pi n_{\mathrm{t}} / \lambda_{\mathrm{c}} $ denotes the wavenumber inside the waveguides with $n_{\mathrm{t}}$ and $\lambda_{\mathrm{c}}$ being the effective refractive index of the waveguide and the wavelength in free space, respectively.
Moreover, according to the equal power model in \cite{wang2025modeling}, we have $\alpha_{i}=1 / \sqrt{M_{\mathrm{t}}} $ for $\forall i$.
The in-waveguide signals will then be emitted at PAs and propagate through a free-space channel, which is given by
\begin{align}
    \mathbf{a}_{\mathrm{b},n}\left( \mathbf{r}_{\mathrm{b}},\mathbf{x}_n \right) =\left[ \frac{\eta e^{-\mathrm{j}k_{\mathrm{c}}r_{\mathrm{b},n,1}}}{r_{\mathrm{b},n,1}},...,\frac{\eta e^{-\mathrm{j}k_{\mathrm{c}}r_{\mathrm{b},n,M_{\mathrm{t}}}}}{r_{\mathrm{b},n,M_{\mathrm{t}}}} \right] ^{\mathrm{H}},
\end{align}
where $\eta \triangleq \frac{\lambda_{\mathrm{c}}}{4 \pi}$ denotes the propagation constant, and $k_{\mathrm{c}}\triangleq 2 \pi / \lambda_{\mathrm{c}}$ denotes the free-space wavenumber.
Furthermore, the received signal \eqref{eq:bob_channel} can be more compactly expressed as
\begin{align}
    y_{\mathrm{b}}=\mathbf{a}_{\mathrm{b}}^{\mathrm{H}}\left( \mathbf{r}_{\mathrm{b}},\mathbf{X} \right) \mathbf{G}\left( \mathbf{X} \right) \mathbf{c}+ n_{\mathrm{b}} = \mathbf{h}_{\rm b}^{\rm H}\mathbf{c} + n_{\mathrm{b}} ,\label{eq:bob_received_signal}
\end{align}
where $\mathbf{X} \triangleq [\mathbf{x}_1, \mathbf{x}_2, ..., \mathbf{x}_{N_{\rm t}}] \in \mathbb{R}^{M_{\mathrm{t}} \times N_{\mathrm{t}}}$ denotes the $x$-coordinate matrix.
Additionally, the overall in-waveguide channel matrix $\mathbf{G}\left( \mathbf{X} \right)\in \mathbb{C} ^{M_{\mathrm{t}}N_{\mathrm{t}}\times N_{\mathrm{t}}}$ and the overall free-space channel vector $\mathbf{a}_{\mathrm{b}}^{}\left( \mathbf{r}_{\mathrm{b}},\mathbf{X} \right)\in \mathbb{C} ^{M_{\mathrm{t}}N_{\mathrm{t}}\times 1}$ are respectively given by
\begin{align}
    &\mathbf{G}\left( \mathbf{X} \right) \triangleq \left[ \begin{matrix}
    	\mathbf{g}_1\left( \mathbf{x}_1 \right)&		\cdots&		\mathbf{0}_{M_{\rm t}}\\
    	\vdots&		\ddots&		\vdots\\
    	\mathbf{0}_{M_{\rm t}}&		\cdots&		\mathbf{g}_{N_{\mathrm{t}}}\left( \mathbf{x}_{N_{\mathrm{t}}} \right)\\
    \end{matrix} \right], \\
    &\mathbf{a}_{\mathrm{b}}^{}\left( \mathbf{r}_{\mathrm{b}},\mathbf{X} \right) \triangleq \left[ \mathbf{a}_{\mathrm{b},1} ^{\mathrm{T}}\left( \mathbf{r}_{\mathrm{b}},\mathbf{x}_1 \right) ,...,\mathbf{a}_{\mathrm{b},N_{\mathrm{t}}} ^{\mathrm{T}}\left( \mathbf{r}_{\mathrm{b}},\mathbf{x}_{N_{\mathrm{t}}} \right) \right] ^{\mathrm{T}}.
\end{align}
where $\mathbf{0}_{M_{\rm t}} \in \mathbb{R}^{M_{\rm t} \times 1}$ denotes the all-zero column vector.

\subsection{Sensing Signal Model} \label{sect:sensing_model}
In this subsection, we model the round‐trip channel between Alice and the adversary Willie, including both the forward probing and the reflected echo.
In what follows, this model will be presented in a downlink-first, uplink-second manner.
\begin{figure}[t!]
    \centering
    \includegraphics[width=0.8\linewidth]{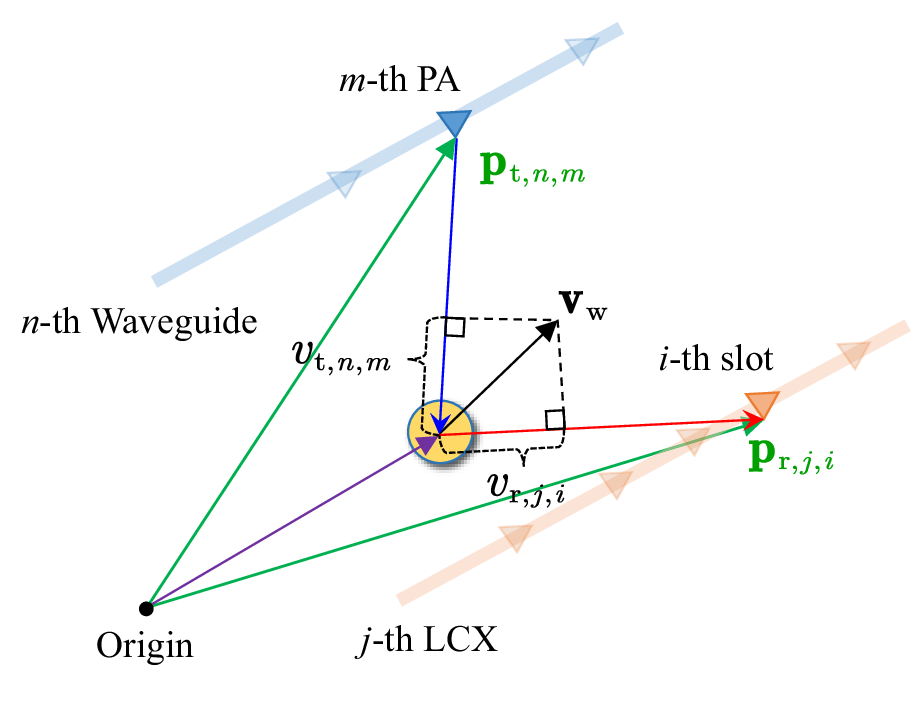}
    \caption{Illustration of the antenna geometry.}
    \label{fig:antenna_geo}
    \vspace{-10pt}
\end{figure}
\subsubsection{Downlink Probing Signal}
Letting the position of Willie be $\mathbf{r}_{\rm w} = [x_{\rm w}, y_{\rm w}, 0]^{\mathrm{T}} \in \mathbb{R}^{3\times 1}$, the distance between the $m$-th PA on the $n$-th waveguide is given by
\begin{align}
    r_{\mathrm{t}, n, m} = \|\mathbf{p}_{\mathrm{t}, n,m} - \mathbf{r}_{\mathrm{w}}\|_2.
\end{align}
Therefore, similar to the modeling techniques used in Section \ref{sect:communication model}, the free-space channel vector from the PAs attached to the $n$-th waveguide $\tilde{\mathbf{a}}_{\mathrm{t},n}\left( \mathbf{r}_{\mathrm{w}},\mathbf{x}_n \right) \in \mathbb{C}^{M_{\rm t} \times 1}$ is given by 
\begin{align}
    \tilde{\mathbf{a}}_{\mathrm{t}, n}\left( \mathbf{r}_{\mathrm{w}},\mathbf{x}_n \right) =\left[ \frac{\eta e^{-\mathrm{j}k_{\mathrm{c}}r_{\mathrm{t},n,1}}}{r_{\mathrm{t},n,1}},...,\frac{\eta e^{-\mathrm{j}k_{\mathrm{c}}r_{\mathrm{t},n,M_{\mathrm{r}}}}}{r_{\mathrm{t},n,M_{\mathrm{r}}}} \right] ^{\mathrm{H}}. \label{eq:free_space_willie}
\end{align}
The phase shifts in \eqref{eq:free_space_willie} are caused by the propagation over distance, referred to as the position-induced phase shift.

In addition, as Willie is moving on the $xy$ plane, the velocity of Willie can be expressed as $\mathbf{v}_{\rm w} = [v_x, v_y, 0]^{\mathrm{T}} \in \mathbb{R}^{3\times 1}$.
Due to Willie's mobility, the Doppler shifts are added to the position-induced phase shift.
More importantly, due to the near-field effects caused by the large aperture size of PASS, the Doppler shifts are not uniform \cite{wang2025near}.
Therefore, the overall velocity $\mathbf{v}_{\rm w}$ needs to be decomposed with respect to different antennas, which is illustrated by Fig. \ref{fig:antenna_geo}.
To model this feature, we define the vector from the PA at $\mathbf{p}_{\mathrm{t}, n,m}$ to Willie as
\begin{align}
    \mathbf{d}_{\mathrm{t}, n, m} \triangleq \mathbf{r}_{\mathrm{w}} - \mathbf{p}_{\mathrm{t}, n,m}. \label{eq:direction_vector}
\end{align}
The direction of $\mathbf{d}_{\mathrm{t}, n, m}$ can be computed by $\hat{\mathbf{d}}_{\mathrm{t}, n,m}=\mathbf{d}_{\mathrm{t}, n,m}/ \|\mathbf{d}_{\mathrm{t}, n,m}\|_2$.
Thus, the projected velocity onto the direction $\hat{\mathbf{d}}_{\mathrm{t}, n,m}$ can be obtained by
\begin{align}
    v_{\mathrm{t}, n,m}&=\left< \hat{\mathbf{d}}_{\mathrm{t}, n,m},\mathbf{v}_{\mathrm{w}} \right> =\mathbf{v}_{\mathrm{w}}^{\mathrm{T}}\hat{\mathbf{d}}_{\mathrm{t}, n, m},
\end{align}
which can be interpreted as the relative velocity to the $m$-th PA on the $n$-th waveguide.
Therefore, considering all PAs on the $n$-th waveguide, the Doppler shift vector $\mathbf{b}_{\mathrm{t},n}^{}\in\mathbb{C}^{M_{\rm t} \times 1}$ is given by
\begin{align}
    \mathbf{b}_{\mathrm{t},n}^{}(\mathbf{v}_{\mathrm{w}},\mathbf{r}_{\mathrm{w}},\mathbf{x}_n)=\left[ e^{-\mathrm{j}k_{\mathrm{c}}\Delta Tv_{\mathrm{t},n,1}},\dots ,e^{-\mathrm{j}k_{\mathrm{c}}\Delta Tv_{\mathrm{t},n,M_{\mathrm{t}}}} \right] ^{\mathrm{H}},
\end{align}
where $\Delta T$ denotes the time duration for one CPI.
Thus, the overall free-space channel vector for the PAs on the $n$-th waveguide can be expressed as
\begin{align}
    \mathbf{a}_{\mathrm{t},n}(\mathbf{v}_{\mathrm{w}},\mathbf{r}_{\mathrm{w}},\mathbf{x}_n) \triangleq \tilde{\mathbf{a}}_{\mathrm{t},n}\left( \mathbf{r}_{\mathrm{w}},\mathbf{x}_n \right) \odot \mathbf{b}_{\mathrm{t},n}^{}(\mathbf{v}_{\mathrm{w}},\mathbf{r}_{\mathrm{w}},\mathbf{x}_n).
\end{align}
Jointly considering all waveguides, the overall probing signal arriving at Willie can be expressed by 
\begin{align}
    y _{\mathrm{w}}&=\sum_{n=1}^{N_{\mathrm{t}}}{\mathbf{a}_{\mathrm{t},n}^{\mathrm{H}}\left( \mathbf{v}_{\mathrm{w}},\mathbf{r}_{\mathrm{w}},\mathbf{x}_n \right) \mathbf{g}_n\left( \mathbf{x}_n \right) c_n}\notag \\
    &=\mathbf{a}_{\mathrm{t}}^{\mathrm{H}}\left( \mathbf{v}_{\mathrm{w}},\mathbf{r}_{\mathrm{w}},\mathbf{X} \right) \mathbf{G}\left( \mathbf{X} \right) \mathbf{c}. \label{eq:received_signal_at_willie}
\end{align}

\subsubsection{Uplink Echo Signal}
The probing signal at Willie will be reflected and captured by the LCXs for sensing.
According to Fig. \ref{fig:antenna_geo}, letting the position of the $i$-th slot on $j$-th LCX be $\mathbf{p}_{\mathrm{r}, j, i}=[x_{\mathrm{r}, j, i}, y_{\mathrm{r}, j, i}, H]^{\mathrm{T}}\in \mathbb{R}^{3 \times 1}$, the distance between $\mathbf{p}_{\mathrm{r}, j, i}$ and $\mathbf{r}_{\mathrm{w}}$ is given by
\begin{align}
    r_{\mathrm{r}, j, i} = \|\mathbf{r}_{\mathrm{w}} - \mathbf{p}_{\mathrm{r}, j,i} \|_2.
\end{align}
The the vector from $\mathbf{p}_{\mathrm{r}, j,i}$ to $\mathbf{r}_{\mathrm{w}}$ can be written as
\begin{align}
    \mathbf{d}_{\mathrm{r},j,i}\triangleq \mathbf{r}_{\mathrm{w}}-\mathbf{p}_{\mathrm{r},j,i},
\end{align}
whose direction can be expressed as $\hat{\mathbf{d}}_{\mathrm{r},j,i}=\mathbf{d}_{\mathrm{r},j,i}/\| \mathbf{d}_{\mathrm{r},j,i}\| _2$.
Given the overall velocity $\mathbf{v}_{\mathrm{w}}$, the relative velocity with respect to the slot at $\mathbf{p}_{\mathrm{r}, j,i}$ can be expressed as
\begin{align}
    v_{\mathrm{r},j,i}=\left< \hat{\mathbf{d}}_{\mathrm{r},j,i},\mathbf{v}_{\mathrm{w}} \right> =\mathbf{v}_{\mathrm{w}}^{\mathrm{T}}\hat{\mathbf{d}}_{\mathrm{r},j,i}.
\end{align}
Considering all slots on the $j$-th LCX, the Doppler shift vector $\mathbf{b}_{\mathrm{r},j}^{} \in \mathbb{C}^{M_{\mathrm{r} \times 1}}$ can be expressed as 
\begin{align}
    \mathbf{b}_{\mathrm{r},j}^{}(\mathbf{v}_{\mathrm{w}}, \mathbf{r}_{\mathrm{w}})=\left[ e^{-\mathrm{j}k_{\mathrm{c}}\Delta Tv_{\mathrm{r},j, 1}},\dots ,e^{-\mathrm{j}k_{\mathrm{c}}\Delta Tv_{\mathrm{r}, j, M_{\mathrm{r}}}} \right] ^{\mathrm{H}}.
\end{align}
Additionally, the distance-induced free-space channel vector for the slots on the $j$-th LCX is given by
\begin{align}
    \tilde{\mathbf{a}}_{\mathrm{r},j}\left( \mathbf{r}_{\mathrm{w}} \right) =\left[ \frac{\eta e^{-\mathrm{j}k_{\mathrm{c}}r_{\mathrm{r},j, 1}}}{r_{\mathrm{r},j, 1}},...,\frac{\eta e^{-\mathrm{j}k_{\mathrm{c}}r_{\mathrm{r},j,M_{\mathrm{r}}}}}{r_{\mathrm{r},j,M_{\mathrm{r}}}} \right] ^{\mathrm{H}}.
\end{align}
Thus, given that the positions of the slots on LCXs are unchanged, the overall free-space channel vector for the $j$-th LCX is captured by
\begin{align}
    \mathbf{a}_{\mathrm{r},j}\left( \mathbf{v}_{\mathrm{w}},\mathbf{r}_{\mathrm{w}} \right) \triangleq \tilde{\mathbf{a}}_{\mathrm{r},j}\left( \mathbf{r}_{\mathrm{w}} \right) \odot \mathbf{b}_{\mathrm{r},j}^{}(\mathbf{v}_{\mathrm{w}},\mathbf{r}_{\mathrm{w}}).
\end{align}
Therefore, the received echo signal on the $j$-th LCX can be expressed as
\begin{align}
    y_{\mathrm{w},j}=\sqrt{\beta} \mathbf{v}_{j}^{\mathrm{T}}\mathbf{a}_{\mathrm{r},j}\left( \mathbf{v}_{\mathrm{w}},\mathbf{r}_{\mathrm{w}} \right) y_{\mathrm{w}} + n_{\mathrm{s},j},
\end{align}
where $n_{\mathrm{s},j} \sim \mathcal{CN}(0, \sigma_{\mathrm{s}}^2)$ denotes the additive Gaussian noise at the $j$-th LCX with power $\sigma_{\mathrm{s}}^2$, $\beta \in \mathbb{C}^{1\times 1}$ denotes the radar-cross section (RCS) of Willie, and the combination vector $\mathbf{v}_j \in \mathbb{C}^{M_\mathrm{r} \times 1}$ is specified by
\begin{align}
    \mathbf{v}_j\triangleq \frac{1}{\sqrt{M_{\mathrm{r}}}}\left[ e^{-\mathrm{j}k_{\mathrm{g}}x_{\mathrm{r},j,1}},...,e_{}^{-\mathrm{j}k_{\mathrm{g}}x_{\mathrm{r},j,M_{\mathrm{r}}}} \right] ^{\mathrm{T}}.
\end{align}
Jointly considering all LCXs and assuming that $n_{\mathrm{s},j}$ for $\forall j$ are independently and identically distributed, the received echo signal can be expressed as
\begin{align}
    \mathbf{y}_{\mathrm{w}}&=\left[ y_{\mathrm{w},1},...,y_{\mathrm{w},N_{\mathrm{r}}} \right] ^{\mathrm{T}}=\beta \mathbf{V}^{\mathrm{T}}\mathbf{a}_{\mathrm{r}}\left( \mathbf{v}_{\mathrm{w}},\mathbf{r}_{\mathrm{w}} \right)y_{\mathrm{w}} +\mathbf{n}_{\mathrm{s}} \notag \\
    &=\sqrt{\beta} \mathbf{V}^{\mathrm{T}}\mathbf{a}_{\mathrm{r}}\left( \mathbf{v}_{\mathrm{w}},\mathbf{r}_{\mathrm{w}} \right) \mathbf{a}_{\mathrm{t}}^{\mathrm{T}}\left( \mathbf{v}_{\mathrm{w}},\mathbf{r}_{\mathrm{w}},\mathbf{X} \right) \mathbf{G}\left( \mathbf{X} \right) \mathbf{c}+\mathbf{n}_{\mathrm{s}}
    \label{eq:measurement_model}
\end{align}
where $\mathbf{n}_{\mathrm{s}} \sim \mathcal{CN}(\boldsymbol{0}_{N_{\mathrm{r}}}, \sigma_{\rm s}^2 \mathbf{I}_{N_{\mathrm{r}}})$ denotes the Gaussian noise, while the combination matrix $\mathbf{V}\in \mathbb{C}^{M_{\mathrm{r}}N_{\mathrm{r}} \times N_{\mathrm{r}}}$ and the free-space channel vector $\mathbf{a}_{\mathrm{r}}\left( \mathbf{v}_{\mathrm{w}},\mathbf{r}_{\mathrm{w}} \right)\in\mathbb{C}^{M_{\mathrm{r}}N_{\mathrm{r}}\times 1}$ are defined by
\begin{align}
\mathbf{V}&\triangleq \left[ \begin{matrix}
	\mathbf{v}_1&		\cdots&		\boldsymbol{0}_{M_{\mathrm{r}}}\\
	\vdots&		\ddots&		\vdots\\
	\boldsymbol{0}_{M_{\mathrm{r}}}&		\cdots&		\mathbf{v}_{N_{\mathrm{r}}}\\
\end{matrix} \right] \notag \\
\mathbf{a}_{\mathrm{r}}\left( \mathbf{v}_{\mathrm{w}},\mathbf{r}_{\mathrm{w}} \right) &\triangleq \left[ \mathbf{a}_{\mathrm{r},1}^{\mathrm{T}}\left( \mathbf{v}_{\mathrm{w}},\mathbf{r}_{\mathrm{w}} \right) ,...,\mathbf{a}_{\mathrm{r},N_{\mathrm{r}}}^{\mathrm{T}}\left( \mathbf{v}_{\mathrm{w}},\mathbf{r}_{\mathrm{w}} \right) \right] ^{\mathrm{T}}.
\end{align}
For notation simplicity, the downlink channel to Willie and the round-trip channel matrix can be respectively defined as
\begin{align}
    \mathbf{h}_{\mathrm{w}}^{\mathrm{H}}\left( \mathbf{v}_{\mathrm{w}},\mathbf{r}_{\mathrm{w}},\mathbf{X} \right) & \triangleq \mathbf{a}_{\mathrm{t}}^{\mathrm{T}}\left( \mathbf{v}_{\mathrm{w}},\mathbf{r}_{\mathrm{w}},\mathbf{X} \right) \mathbf{G}\left( \mathbf{X} \right),  \label{eq:willie_cis}\\
     \mathbf{H}_{\mathrm{w}}\left( \mathbf{v}_{\mathrm{w}},\mathbf{r}_{\mathrm{w}},\mathbf{X} \right) &\triangleq \beta \mathbf{V}^{\mathrm{T}}\mathbf{a}_{\mathrm{r}}^{}\left( \mathbf{v}_{\mathrm{w}},\mathbf{r}_{\mathrm{w}} \right) \mathbf{a}_{\mathrm{t}}^{\mathrm{T}}\left( \mathbf{v}_{\mathrm{w}},\mathbf{r}_{\mathrm{w}},\mathbf{X} \right) \mathbf{G}\left( \mathbf{X} \right) \label{eq:define_H}.
\end{align}

\section{Sensing-Assisted Covert Communication Empowered by PASS} \label{sect:malicious_tracking}
In this section, we begin by analyzing the covertness of the proposed system and deriving a corresponding metric that is parameterized by Willie’s CSI. 
To obtain Willie’s CSI, we employ an EKF approach to track his trajectory in real time. 
Building on these results, we then formulate the covert communication problem.

\subsection{Covertness Analysis}
As the malicious user, Willie aims to detect whether Alice is transmitting confidential signals to Bob.
In the pursuit of transmission covertness, Alice will insert AN into the transmit signals to misguide Willie.
Specifically, the transmit signal at Alice $\mathbf{s}$ can be expressed as
\begin{align}
    \mathbf{c} = \mathbf{w}s + \mathbf{q},
\end{align}
where $s$ denotes the unit-power intended signal for Bob, $\mathbf{w}\in \mathbb{C}^{N_{\mathrm{t}} \times 1}$ denotes the baseband beamforming vector at PASS, and $\mathbf{q}\in\mathbb{C}^{N_{\rm t} \times 1}$ denotes the AN vector.
Then, according to \eqref{eq:received_signal_at_willie}, the received signal at Willie can be expressed as 
\begin{align}
    y_{\mathrm{w}}=\mathbf{a}_{\mathrm{t}}^{\mathrm{T}}\left( \mathbf{v}_{\mathrm{w}},\mathbf{r}_{\mathrm{w}},\mathbf{X} \right) \mathbf{G}\left( \mathbf{X} \right) \mathbf{s}+n_{\mathrm{w}}=\mathbf{h}_{\mathrm{w}}^{\mathrm{H}}\left( \mathbf{v}_{\mathrm{w}},\mathbf{r}_{\mathrm{w}},\mathbf{X} \right) \mathbf{s}+n_{\mathrm{w}},
\end{align}
where $n_{\mathrm{w}} \sim \mathcal{CN}(0, \sigma_{\rm w}^2)$ denotes the additive complex-valued Gaussian noise with zero mean and a power of $\sigma_{\rm w}^2$.
Willie has two hypotheses on the transmission behavior at Alice, i.e., $\mathcal{H}_0$: Alice is silent, and $\mathcal{H}_1$: Alice is transmitting signals to Bob.
Under the two hypotheses $\mathcal{H}_0$ and $\mathcal{H}_1$, the likelihood functions can be respectively specified by 
\begin{align}
    \left\{ \begin{matrix}
	p_{\left| y_{\mathrm{w}} \right|^2,0}\left( x \right) =\frac{1}{\pi \lambda _0}e^{-\frac{x}{\lambda _0}},&		\mathcal{H} _0,\\
	p_{\left| y_{\mathrm{w}} \right|^2,1}\left( x \right) =\frac{1}{\pi \lambda _1}e^{-\frac{x}{\lambda _1}},&		\mathcal{H} _1,\\
\end{matrix} \right. 
\end{align}
where $\lambda _0=\left| \mathbf{h}_{\mathrm{w}}^{\mathrm{H}}\left( \mathbf{v}_{\mathrm{w}},\mathbf{r}_{\mathrm{w}},\mathbf{X} \right) \mathbf{q} \right|^2+\sigma _{\rm w}^{2}$ and $\lambda _1=\left| \mathbf{h}_{\mathrm{w}}^{\mathrm{H}}\left( \mathbf{v}_{\mathrm{w}},\mathbf{r}_{\mathrm{w}},\mathbf{X} \right) \mathbf{q} \right|^2+\left| \mathbf{h}_{\mathrm{w}}^{\mathrm{H}}\left( \mathbf{v}_{\mathrm{w}},\mathbf{r}_{\mathrm{w}},\mathbf{X} \right) \mathbf{w} \right|^2+\sigma _{\rm w}^{2}$.
Under the assumption that each hypothesis becomes true with an equal probability \cite{bash2013limits, wang2024sensing}, the binary decisions are made according to the following rule:
\begin{align}
    \frac{p_{\left| y_{\mathrm{w}} \right|^2,1}\left( x \right)}{p_{\left| y_{\mathrm{w}} \right|^2,0}\left( x \right)}
\mathrel{\substack{\mathcal{D}_1 \\ \gtrless \\ \mathcal{D}_0}} 1 \Rightarrow |y_{\mathrm{w}}|^2
\mathrel{\substack{\mathcal{D}_1 \\ \gtrless \\ \mathcal{D}_0}} \frac{\lambda _0\lambda _1}{\lambda _1-\lambda _0}\ln \frac{\lambda _1}{\lambda _0}.
\end{align}
According to the above decision rule, the total detection error probability, including both miss detection and false alarm, can be expressed as \cite{bash2013limits, wang2024sensing}:
\begin{align}
    \rho (\mathbf{q}, \mathbf{w}, \mathbf{X}) = 1 - V_T(p_{\left| y_{\mathrm{w}} \right|^2,0}\left( x \right), p_{\left| y_{\mathrm{w}} \right|^2,1}\left( x \right)),
\end{align}
where $V_T(p_{\left| y_{\mathrm{w}} \right|^2,0}\left( x \right), p_{\left| y_{\mathrm{w}} \right|^2,1}\left( x \right)) \triangleq \frac{1}{2}\|p_{\left| y_{\mathrm{w}} \right|^2,0}\left( x \right) - p_{\left| y_{\mathrm{w}} \right|^2,1}\left( x \right)\|_1$ is the total variation distance between distributions $p_{\left| y_{\mathrm{w}} \right|^2,0}\left( x \right)$ and $p_{\left| y_{\mathrm{w}} \right|^2,1}\left( x \right)$.
To make this expression more tractable, Pinsker's inequality \cite{cover2006elements} can be utilized to characterize the upper bound of $V_T(p_{\left| y_{\mathrm{w}} \right|^2,0}\left( x \right), p_{\left| y_{\mathrm{w}} \right|^2,1}\left( x \right))$, which can be derived as
\begin{align}
    &V_T(p_{\left| y_{\mathrm{w}} \right|^2,0}\left( x \right),p_{\left| y_{\mathrm{w}} \right|^2,1}\left( x \right)) \notag \\ 
    &\qquad \qquad \le \sqrt{0.5 D_{\mathrm{KL}}( p_{\left| y_{\mathrm{w}} \right|^2,0}\left( x \right) \Vert p_{\left| y_{\mathrm{w}} \right|^2,1}\left( x \right))},
\end{align}
where $D_{\mathrm{KL}}( p_{\left| y_{\mathrm{w}} \right|^2,0}\left( x \right)\Vert p_{\left| y_{\mathrm{w}} \right|^2,1}\left( x \right))$ denotes the KL divergence between distributions $ p_{\left| y_{\mathrm{w}} \right|^2,0}\left( x \right)$ and $ p_{\left| y_{\mathrm{w}} \right|^2,1}\left( x \right)$.
Its detailed expressions can be derived in a closed form: 
\begin{align}
    &D_{\mathrm{KL}}( p_{\left| y_{\mathrm{w}} \right|^2,0}\left( x \right) \Vert p_{\left| y_{\mathrm{w}} \right|^2,1}\left( x \right) ) \notag \\
    &=\int_{-\infty}^{+\infty}{p_{\left| y_{\mathrm{w}} \right|^2,0}\left( x \right)\ln \frac{p_{\left| y_{\mathrm{w}} \right|^2,0}\left( x \right)}{p_{\left| y_{\mathrm{w}} \right|^2,1}\left( x \right)}}dx=\ln \frac{\lambda _1}{\lambda _0}+\frac{\lambda _0}{\lambda _1}-1.
\end{align}
To fulfill the covertness requirement, we have the following criteria:
\begin{align}
    D_{\mathrm{KL}}( p_{\left| y_{\mathrm{w}} \right|^2,0}\left( x \right) \Vert p_{\left| y_{\mathrm{w}} \right|^2,1}\left( x \right)) \le 2\epsilon^2, \label{eq:covertness_KL}
\end{align}
where $\epsilon\in[0, 1)$ denoted the covertness threshold.
Therefore, a metric measuring the transmission covertness is presented by the left-hand side of inequality \eqref{eq:covertness_KL}.
In particular,  covert constraint \eqref{eq:covertness_KL} denotes the upper bound of the total error rate at the malicious user.
By choosing a smaller threshold $\epsilon$, Alice can achieve a higher level of transmission covertness, since the adversary’s error rate will correspondingly increase.
However, the covert constraint \eqref{eq:covertness_KL} is parameterized by the CSI of the malicious user, which is typically inaccessible to Alice.
Therefore, the sensing functionality of PASS is utilized to track the adversary by adopting EKF, which will be detailed in the subsequent subsection.

\subsection{EKF-Based Adversary Tracking Scheme}
To obtain the CSI of Willie, we present an EKF approach for tracking Willie's movement. 
Before proceeding, we first present the kinematic model, which describes the evolution rule of Willie's mobility status.
Specifically, this model can be expressed as
\begin{equation}
     \begin{cases}
	v_{x,t}^{}=v_{x,t-1}^{}+\Delta v_{x}^{},\\
	v_{y,t}^{}=v_{y,t-1}^{}+\Delta v_{y}^{},\\
\end{cases}
\begin{cases}
	x_{\mathrm{w},t}^{}=x_{\mathrm{w},t-1}^{}+v_{x,t-1}^{}\Delta T,\\
	y_{\mathrm{w},t}^{}=y_{\mathrm{w},t-1}^{}+v_{y,t-1}^{}\Delta T,\\
\end{cases}\label{eq:kinematic_equations}
\end{equation}
where $\Delta v_{x} \in \mathcal{N}(0, \sigma_{v_x}^2)$ and $\Delta v_{y} \in \mathcal{N}(0, \sigma_{v_y}^2)$ denote the velocity variances within one CPI along the $x$- and $y$-axes with $\sigma_{v_x}^2$ and $\sigma_{v_y}^2$ being the deviations of the Gaussian distributions.
In the following sub-sections, we will first present the state transition and observation models for the EKF.
Building on this, the EKF framework is explained, which consists of two main steps: prior prediction and posterior update. 

\subsubsection{State Transition and Observation Models}
According to the kinematic model, $\boldsymbol{\xi }_t=[x_{\mathrm{w},t},y_{\mathrm{w},t},v_{x,t}^{},v_{y,t}^{}]^{\mathrm{T}}\in \mathbb{R} ^{4\times 1}$ denotes mobility status of Willie during the $(t-1)$-th CPI.
Therefore, building on \eqref{eq:kinematic_equations}, the state transition model from CPI $t-1$ to CPI $t$ can be compactly expressed by 
\begin{align}
    \boldsymbol{\xi}_{t} = g(\boldsymbol{\xi}_{t-1}) + \boldsymbol{\delta}_{t-1}, \label{eq:state transition model}
\end{align}
where $g(\cdot)$ denotes the states transition matrix given by 
\begin{align}
    g\left( \boldsymbol{\xi }_{t} \right) =\left[ \begin{matrix}
	1&		0&		\Delta T&		0\\
	0&		1&		0&		\Delta T\\
	0&		0&		1&		0\\
	0&		0&		0&		1\\
\end{matrix} \right] \boldsymbol{\xi }_{t-1}. \label{eq:kinematic_model}
\end{align}
In addition, $\boldsymbol{\delta}_{t-1} \sim \mathcal{N}(\boldsymbol{0}_4, \mathbf{Q}_{\boldsymbol{\xi}})$ denotes the motion noise vector during the $(t-1)$-th CPI, and the covariance matrix is specified by $\mathbf{Q}_{\boldsymbol{\xi}} = \mathrm{diag}\{0,0, \sigma^2_{v_x},\sigma^2_{v_y}\} \in \mathbb{R}^{4 \times 4}$.

The observation model describes the modeling of the received echoes at Alice.
Thereby, according to Section \ref{sect:sensing_model}, the observation model of Willie at the $t$-th CPI can be expressed as
\begin{align}
    \boldsymbol{\phi }_{\mathrm{w}, t-1}=h\left( \boldsymbol{\xi }_{t-1} \right) +\mathbf{n}_{t-1}, \label{eq:observation model}
\end{align}
where $h\left( \cdot\right)$ represents the overall modeling of the round-trip channel, that is detailed in \eqref{eq:measurement_model}.
Moreover, the noise term can be expressed as $\mathbf{n}_{t-1}\sim \mathcal{C} \mathcal{N} (\boldsymbol{0}_{N_{\rm r}}, \mathbf{R})$ with $ \mathbf{R} = \sigma_{\mathrm{s}}^2 \mathbf{I}_{N_{\rm r}}$ being the covariance matrix.

\subsubsection{Prior Prediction}
In this step, the objective is to predict the next mobility status $\boldsymbol{\xi}_{t}$ based on the current mobility state $\boldsymbol{\xi}_{t-1}$ though the state transition model captured by \eqref{eq:state transition model}.
Specifically, the current mobility status will be fed to $g(\cdot)$ to form the prior prediction on the mobility status $\hat{\boldsymbol{\xi}}_{t}$, i.e.,
\begin{align}
    \hat{\boldsymbol{\xi}}_{t} = g(\boldsymbol{\xi}_{t-1}), \label{eq:prior_mobility}
\end{align}
Simultaneously, the covariance estimation of Willie can be updated according to
\begin{align}
    \mathbf{P}_{t \mid t-1}=\mathbf{G}_{t-1}^{}\mathbf{P}_{t-1}^{}\mathbf{G}_{t-1}^{\mathrm{H}}+\mathbf{Q}_{\boldsymbol{\xi}}. \label{eq:coverianxce_prior}
\end{align}
where $\mathbf{G}_{t-1}$ denotes the Jacobian matrix of the kinematic model in \eqref{eq:kinematic_model}.
At this point, the prior estimation of the mobility status and the covariance estimation are obtained.
In the sequel, the prior predictions are updated according to the posterior observations, i.e., received echo signals.

\subsubsection{Posterior Update}
After the echo signal in the $t$-th CPI is received, we rectify our prior estimate of Willie's mobility status.
In the first place, we compute the expected observations using our prior estimation $\hat{\boldsymbol{\xi}}$ via 
\begin{align}
    \hat{\mathbf{y}}_{\mathrm{w}, t}=h( \hat{\boldsymbol{\xi}}_{t}).
\end{align}
Then, the residual term between the real observations $\boldsymbol{\xi}_{t}$ and the synthesized observation $\hat{\boldsymbol{\xi}}_{t}$ is quantified by 
\begin{align}
    \mathbf{u}_{t}=\mathbf{y}_{\mathrm{w}, t}-\hat{\mathbf{y}}_{\mathrm{w}, t},
\end{align}
which measures the unexpected part of the prior estimation.
To update the EKF with the $\mathbf{u}_{t}$, the linearized version of the observation model $h(\cdot)$ is computed at point $\hat{\boldsymbol{\xi}}_{t}$ by taking partial derivatives, i.e.,
\begin{align}
    \mathbf{J}_{t \mid t-1}=\frac{\partial h\left( \boldsymbol{\xi } \right)}{\partial \boldsymbol{\xi }}\mid_{\boldsymbol{\xi }=\hat{\boldsymbol{\xi}}_{t}}^{}, \label{eq:linearization observation model}
\end{align}
where $\mathbf{J}_{t \mid t-1}$ denotes the Jacobian matrix of the observation model with respect to the mobility status.
The residual covariance matrix can be calculated according to 
\begin{align}
\mathbf{S}_{t}=\mathbf{J}_{t \mid t-1}\mathbf{P}_{t \mid t-1}\mathbf{J}_{t \mid t-1}^{\mathrm{H}}+\mathbf{R}. \label{eq:S}
\end{align}
Hence, the Kalman gain is given by 
\begin{align}
    \mathbf{K}_{t}=\mathbf{P}_{t \mid t-1}\mathbf{J}_{t \mid t-1}^{\mathrm{H}}\mathbf{S}_{ t}^{-1}. \label{eq:kalman_gain}
\end{align}
Based on \eqref{eq:kalman_gain}, the mobility status vector can be updated in a posterior fashion via
\begin{align}
    \boldsymbol{\xi }_{t}=\hat{\boldsymbol{\xi}}_{t}+\mathbf{K}_{t}\mathbf{u}_{t}. \label{eq:update_mobility}
\end{align}
Finally, the covariance matrix is updated according to
\begin{align}
    \mathbf{P}_{t}=\left( \mathbf{I}_{4}-\mathbf{K}_{t}\mathbf{J}_{t\mid t-1} \right) \mathbf{P}_{t\mid t-1}.\label{eq:covariance_update}
\end{align}
To maintain consistency in the logit flow, the detailed derivation of \eqref{eq:linearization observation model} is provided in Appendix \ref{appendix:jacobian_of_observation}.

\subsection{Problem Formulation}
According to the results obtained in the previous section, we can track Willie's mobility status using the prior estimation $\hat{\boldsymbol{\xi}}_t$.
Building on this, in this section, we will first derive the communication and sensing metric for the ISAC system, followed by a covert communication problem formulation.
Then, the solution to the formulated optimization problem is presented.

Considering to \eqref{eq:bob_received_signal}, the SINR at Bob is given by
\begin{align}
    \gamma_t \left( \mathbf{w}_t,\mathbf{q}_t,\mathbf{X}_t \right) =\frac{\left| \mathbf{h}_{\mathrm{b}}^{\mathrm{H}}\left( \mathbf{r}_{\mathrm{b}},\mathbf{X}_t \right) \mathbf{w}_t \right|^2}{\left| \mathbf{h}_{\mathrm{b}}^{\mathrm{H}}\left( \mathbf{r}_{\mathrm{b}},\mathbf{X}_t \right) \mathbf{q}_t \right|^2+\sigma _{\rm b}^{2}}.
\end{align}
Given the total number of CPI is $T$, the average covert rate at Bob can be expressed as 
\begin{align}
    R\left( \mathcal{W} ,\mathcal{Q} ,\mathcal{X} \right) =\mathbb{E} _t\left\{ \log _2\left( 1+\gamma _t\left( \mathbf{w}_t,\mathbf{q}_t,\mathbf{X}_t \right) \right) \right\} ,
\end{align}
where $\mathcal{W} \triangleq \left\{ \mathbf{w}_t \right\} _{t=1}^{T}$, $
\mathcal{Q} \triangleq \left\{ \mathbf{q}_t \right\} _{t=1}^{T}$, and $\mathcal{X} \triangleq \left\{ \mathbf{X}_t \right\} _{t=1}^{T}$.
For the sensing functionality, we need to ensure the power of the received echo signal exceeds a predefined threshold, i.e., $\Gamma_{\rm sen}$.
Therefore, the sensing constraint for the $t$-th CPI can be expressed as 
\begin{align}
    &g(\mathbf{c}_t,\mathbf{X}_t)\triangleq \notag \\&\mathbf{c}_{t}^{\mathrm{H}}\mathbf{H}_{\mathrm{w}}^{\mathrm{H}}\left( {\mathbf{v}}_{\mathrm{w},t},{\mathbf{r}}_{\mathrm{w},t},\mathbf{X}_t \right) \mathbf{H}_{\mathrm{w}}\left( {\mathbf{v}}_{\mathrm{w},t},{\mathbf{r}}_{\mathrm{w},t},\mathbf{X}_t \right) \mathbf{c}_{t}^{}\ge \Gamma _{\mathrm{sen}}. \label{eq:sensing}
\end{align}
Note that due to the non-cooperative behavior of Willie, the real-time perfect CSI $\mathbf{H}_{\mathrm{w}}^{}\left( \mathbf{v}_{\mathrm{w},t},\mathbf{r}_{\mathrm{w},t},\mathbf{X}_t \right) $ cannot be obtained. 
However, with the prior mobility status estimation $\hat{\boldsymbol{\xi}}_t=[\hat{\mathbf{v}}_{{\rm w}, t}^{\rm T}, \hat{\mathbf{r}}_{{\rm w}, t}^{\rm T}]^{\rm T}$, the estimated version of $\mathbf{H}_{\mathrm{w}}^{}\left( \mathbf{v}_{\mathrm{w},t},\mathbf{r}_{\mathrm{w},t},\mathbf{X}_t \right) $ denoted by $\mathbf{H}_{\mathrm{w}}^{}\left( \hat{\mathbf{v}}_{\mathrm{w},t},\hat{\mathbf{r}}_{\mathrm{w},t},\mathbf{X}_t \right) $ can be constructed to compensate for the absence of instantaneous CSI.

Therefore, the optimization problem can be formulated as 
\begin{subequations}
\label{problem:original}
\begin{align}
  \max_{\mathcal{W},\mathcal{Q},\mathcal{X}}\quad
    & R\left( \mathcal{W} ,\mathcal{Q} ,\mathcal{X}\right) \label{objective}\\[6pt]
  \text{s.t.}\quad
    & D_{\mathrm{KL}}(p_{|y_{\mathrm{w}}|^2,0}(x)\,\|\,p_{|y_{\mathrm{w}}|^2,1}(x))=0, \label{constraint:corvertness} \\
    & g(\mathbf{c}_t, \mathbf{X}_t) \ge \Gamma_{\mathrm{sen}}, ~\forall t, \label{constraint:sensing}\\
    & \|\mathbf{w}_t\|^2_2 + \|\mathbf{q}_t\|^2_2 = P_{\max}, ~\forall t, \label{constraint:tot_power} \\
    & 0\leq x_{n,m,t} \leq L_{\max},~\forall m,n,t, \label{constraint:position_1}\\
    &\Delta_{n,m,t}^p\triangleq x_{n,m,t}-x_{n,m-1,t}\geq \Delta_{\min},~\forall m,n,t. \label{constraint:position_2}
\end{align}
\end{subequations}
The objective \eqref{objective} is to maximize the average covert rate for the legal user, Bob.
For the constraints, the constraint \eqref{constraint:corvertness} is the perfect covertness requirement, achieved at $\epsilon=0$ in \eqref{eq:covertness_KL}.
Equivalently, by adding \eqref{constraint:corvertness}, there will be no information leakage to the illegal user, Willie.
Constraint \eqref{constraint:sensing} denotes the sensing requirement, ensuring that the tracking results are accurate.
Moreover, the constraint \eqref{constraint:tot_power} confines the transmit power according to the power budget $P_{\max}$.
Additionally, by the format of \eqref{constraint:tot_power}, we assume that the beamforming vector and the AN signal are independent.
Finally, the constraints \eqref{constraint:position_1} and \eqref{constraint:position_2} ensure the position of PAs within the feasible position region, and the spacings between adjacent PAs are sufficiently large to avoid mutual coupling.

\section{Reinforcement Learning-Based Solution} \label{sect:solution}
The optimization problem \eqref{problem:original} can be divided into three subproblems, concerning $\mathcal{W}$, $\mathcal{Q}$, and $\mathcal{X}$, respectively.
Irrespective of the optimization of PA's positions $\mathcal{X}$, the problem \eqref{problem:original} can be effectively solved by the semi-definite relaxation (SDR) method.
Then, when the optimization of $\mathcal{X}$ is considered, we have to resort to an element-wise algorithm with SDR to solve \eqref{problem:original}.
In particular, for any $\mathbf{X}_t \in \mathcal{X}$, an SDR algorithm needs to be employed, thereby leading to a complexity of $\mathcal{O}(KN_{\rm t}^{3.5})$ \cite{luo2010semi}, where $K=M_{\rm t} N_{\rm t}$ denotes the total number of PAs.
This complexity is impractical for the proposed tracking system.
Therefore, to address this issue, we first drive the optimal solutions of the beamforming vector $\mathbf{w}_t$ and the AN signal $\mathbf{q}_t$, for any given CPI $t$.
Accordingly, the optimal beamforming vectors and AN signals across all CPIs are given by $\mathcal{W}=\{\mathbf{w}_1, ..., \mathbf{w}_T\}$ and $\mathcal{Q}=\{\mathbf{q}_1, ..., \mathbf{q}_T\}$, respectively.
However, the rest sub-problem with respect to $\mathcal{X}$ does not have an optimal solution and therefore cannot be solved in the same manner.
To address this issue, we employ a DRL-based approach, as Willie's trajectory is temporally correlated.

\subsubsection{Subproblem with respect to $\mathcal{W}$} \label{sect:subproblem_wrt_W}
According to the above discussions, any element $\mathbf{w}_t$ in $\mathcal{W}$ can be analyzed individually.
Thus, for notional simplicity, we drop the index of CPI.
Fixed the rest optimization variables, the subproblem with respect to $\mathbf{w}$ can be formulated as
\begin{subequations}
\label{subproblem_w}
\begin{align}
  \max_{\mathbf{w}}\quad
    & \left| \mathbf{h}_{\mathrm{b}}^{\mathrm{H}}\left( \mathbf{r}_{\mathrm{b}},\mathbf{X} \right) \mathbf{w} \right|^2
 \label{objective_commu}\\[6pt]
  \text{s.t.}\quad
    &\eqref{constraint:corvertness}, \eqref{constraint:sensing}, \text{ and }\eqref{constraint:tot_power} \notag
\end{align}
\end{subequations}
First, we examine the covertness constraint in \eqref{constraint:corvertness} and delineate the conditions under which covertness is ensured.
For the covertness constraint \eqref{constraint:corvertness}, we have the following derivations:
\begin{align}
   & \ln \frac{\lambda _1}{\lambda _0}+\frac{\lambda _0}{\lambda _1}-1=0\Rightarrow \lambda _1=\lambda _0 \notag \\
&\Rightarrow \lambda _0=\left| \mathbf{h}_{\mathrm{w}}^{\mathrm{H}}\left( \hat{\mathbf{v}}_{\mathrm{w}},\hat{\mathbf{r}}_{\mathrm{w}},\mathbf{X} \right) \mathbf{q} \right|^2+\sigma _{\mathrm{w}}^{2} \notag \\
&=\lambda _1=\left| \mathbf{h}_{\mathrm{w}}^{\mathrm{H}}\left( \hat{\mathbf{v}}_{\mathrm{w}},\hat{\mathbf{r}}_{\mathrm{w}},\mathbf{X} \right) \mathbf{q} \right|^2+\left| \mathbf{h}_{\mathrm{w}}^{\mathrm{H}}\left( \hat{\mathbf{v}}_{\mathrm{w}},\hat{\mathbf{r}}_{\mathrm{w}},\mathbf{X} \right) \mathbf{w} \right|^2+\sigma _{\mathrm{w}}^{2} \notag \\
&\Rightarrow \left| \mathbf{h}_{\mathrm{w}}^{\mathrm{H}}\left( \hat{\mathbf{v}}_{\mathrm{w}},\hat{\mathbf{r}}_{\mathrm{w}},\mathbf{X} \right) \mathbf{w} \right|^2=0,
\end{align}
In the above mathematical manipulations, we utilize the estimated mobility status to deal with the unknown CSI issue at Willie.
Therefore, we have $\mathbf{h}_{\mathrm{w}} \left( \hat{\mathbf{v}}_{\mathrm{w}},\hat{\mathbf{r}}_{\mathrm{w}},\mathbf{X} \right) \perp \mathbf{w} $.
Building on this property, we further have $\mathbf{H}_{\mathrm{w}}^{}\left( \hat{\mathbf{v}}_{\mathrm{w}},\hat{\mathbf{r}}_{\mathrm{w}},\mathbf{X} \right) \mathbf{w}=\boldsymbol{0}_{N_{\rm r}}$, as $\mathbf{h}_{\mathrm{w}} \left( \hat{\mathbf{v}}_{\mathrm{w}},\hat{\mathbf{r}}_{\mathrm{w}},\mathbf{X} \right)$ is the transmit response of $\mathbf{H}_{\mathrm{w}}^{}\left( \hat{\mathbf{v}}_{\mathrm{w}},\hat{\mathbf{r}}_{\mathrm{w}},\mathbf{X} \right)$ according to \eqref{eq:define_H}.
Hence, the sensing constraint \eqref{constraint:sensing} can be simplified and reformulated by 
\begin{align}
    g(\mathbf{c},\mathbf{X}) &\triangleq \mathbf{c}^{\mathrm{H}}\mathbf{H}_{\mathrm{w}}^{\mathrm{H}}\left( \hat{\mathbf{v}}_{\mathrm{w}},\hat{\mathbf{r}}_{\mathrm{w}},\mathbf{X} \right) \mathbf{H}_{\mathrm{w}}^{}\left( \hat{\mathbf{v}}_{\mathrm{w}},\hat{\mathbf{r}}_{\mathrm{w}},\mathbf{X} \right) \mathbf{c}  \notag \\
    &= \mathbf{q}^{\mathrm{H}}\mathbf{H}_{\mathrm{w}}^{\mathrm{H}}\left( \hat{\mathbf{v}}_{\mathrm{w}},\hat{\mathbf{r}}_{\mathrm{w}},\mathbf{X} \right) \mathbf{H}_{\mathrm{w}}^{}\left( \hat{\mathbf{v}}_{\mathrm{w}},\hat{\mathbf{r}}_{\mathrm{w}},\mathbf{X} \right) \mathbf{q} \notag \\
    &=g(\mathbf{q},\mathbf{X})\ge \Gamma _{\mathrm{sen}}
    , \label{eq:constraint_sensing_new}
\end{align}
which is irrelevant to $\mathbf{w}$ and parameterized solely by $\mathbf{q}$.
According to the objective function \eqref{objective_commu}, the direction of the optimal $\mathbf{w}$ is to align $\mathbf{w}$ with Bob's channel $\mathbf{h}_{\mathrm{b}}$ while being orthogonal to Willie's channel $\mathbf{h}_{\mathrm{w}}^{}\left( \hat{\mathbf{v}}_{\mathrm{w}},\hat{\mathbf{r}}_{\mathrm{w}},\mathbf{X} \right) $.
Therefore, the optimal solution can be expressed as
\begin{align}
    \mathbf{w}^{\star}=\frac{\sqrt{P_{\rm t}}\left( \mathbf{I}_{N_{\rm t}}-\frac{\mathbf{h}_{\mathrm{w}}^{}\left( \hat{\mathbf{v}}_{\mathrm{w}},\hat{\mathbf{r}}_{\mathrm{w}},\mathbf{X} \right) \mathbf{h}_{\mathrm{w}}^{\mathrm{H}}\left( \hat{\mathbf{v}}_{\mathrm{w}},\hat{\mathbf{r}}_{\mathrm{w}},\mathbf{X} \right)}{\left\| \mathbf{h}_{\mathrm{w}}^{}\left( \hat{\mathbf{v}}_{\mathrm{w}},\hat{\mathbf{r}}_{\mathrm{w}},\mathbf{X} \right) \right\| _{2}^{2}} \right) \mathbf{h}_{\mathrm{b}}^{}\left( \mathbf{r}_{\mathrm{b}},\mathbf{X} \right)}{\left\| \left( \mathbf{I}_{N_{\rm t}}-\frac{\mathbf{h}_{\mathrm{w}}^{}\left( \hat{\mathbf{v}}_{\mathrm{w}},\hat{\mathbf{r}}_{\mathrm{w}},\mathbf{X} \right) \mathbf{h}_{\mathrm{w}}^{\mathrm{H}}\left( \hat{\mathbf{v}}_{\mathrm{w}},\hat{\mathbf{r}}_{\mathrm{w}},\mathbf{X} \right)}{\left\| \mathbf{h}_{\mathrm{w}}^{}\left( \hat{\mathbf{v}}_{\mathrm{w}},\hat{\mathbf{r}}_{\mathrm{w}},\mathbf{X} \right) \right\| _{2}^{2}} \right) \mathbf{h}_{\mathrm{b}}^{}\left( \mathbf{r}_{\mathrm{b}},\mathbf{X} \right) \right\| _2}, \label{eq:solution_to_w}
\end{align}
where $P_{\rm t}>0$ denotes the optimal power allocated for the beamforming design.

\subsubsection{Subproblem with respect to $\mathcal{Q}$} \label{sect:subproblem_to_Q}
The remaining problem is to design the AN vector set $\mathcal{Q}$.
Likewise, we analyze an arbitrary CPI $\mathbf{q}_t\in\mathcal{Q}$, and drop CPI's index $t$ for brevity.
For simplicity, we write $\mathbf{q} = \sqrt{P}\tilde{\mathbf{q}}$, where $P \ge 0$ denotes the allocated power for $\mathbf{q}$ and $\tilde{\mathbf{q}}$ is the unit length vector, satisfying $\|\tilde{\mathbf{q}}\|_2 = 1$.
For this subproblem, the problem can be reformulated as 
\begin{subequations}
\label{subproblem_w}
\begin{align}
  \max_{\mathbf{q}} \quad &\frac{\left(P_{\max} - P \right) \left| \mathbf{h}_{\mathrm{b}}^{\mathrm{H}}\left( \mathbf{r}_{\mathrm{b}},\mathbf{X} \right) \tilde{\mathbf{w}} \right|^2}{P\left| \mathbf{h}_{\mathrm{b}}^{\mathrm{H}}\left( \mathbf{r}_{\mathrm{b}},\mathbf{X} \right) \tilde{\mathbf{q}} \right|+\sigma _{\rm b}^{2}}
\label{objective_sensing}\\[6pt]
  \text{s.t.}\quad
    &Pg(\tilde{\mathbf{q}},\mathbf{X})\ge \Gamma _{\mathrm{sen}}. \label{constraint:new_sensing}
\end{align}
\end{subequations}
Using proof by contradiction, it is easy to prove that the optimal $P$ is obtained when the constraint \eqref{constraint:new_sensing} reaches its boundary, i.e., $P^{\star}=\Gamma _{\mathrm{sen}}/g(\tilde{\mathbf{q}},\mathbf{X})$.
Under these circumstances, the inequality constraint can be replaced by an equivalent equality constraint.
Then, considering $\Gamma_{\rm sen} > 0$, by plugging $P^{\star}$ into the objective function, the problem \eqref{subproblem_w} can be rewritten as 
\begin{subequations}
\label{subproblem_w_sub_1}
\begin{align}
  \max_{\mathbf{q}} \quad &c\left( \mathbf{X} \right) \frac{\tilde{\mathbf{q}}^{\mathrm{H}}\mathbf{A}\left( \mathbf{X} \right) \tilde{\mathbf{q}}}{\tilde{\mathbf{q}}^{\mathrm{H}}\mathbf{B}\left( \mathbf{X} \right) \tilde{\mathbf{q}}}.
\label{objective_sensing_w_sub1}
\end{align}
\end{subequations}
where the matrices $\mathbf{A}\left( \mathbf{X} \right)$ and $\mathbf{B}\left( \mathbf{X} \right)$ and the constant are defined as
\begin{align}
    \mathbf{A}\left( \mathbf{X} \right) &\triangleq P_{\max}\mathbf{H}_{\mathrm{w}}^{\mathrm{H}}\left( \hat{\mathbf{v}}_{\mathrm{w}},\hat{\mathbf{r}}_{\mathrm{w}},\mathbf{X} \right) \mathbf{H}_{\mathrm{w}}^{}\left( \hat{\mathbf{v}}_{\mathrm{w}},\hat{\mathbf{r}}_{\mathrm{w}},\mathbf{X} \right) -\Gamma _{\mathrm{sen}}\mathbf{I}_{N_{\mathrm{r}}}, \\
    \mathbf{B}\left( \mathbf{X} \right) &\triangleq \Gamma _{\mathrm{sen}}\mathbf{h}_{\mathrm{b}}^{}\left( \mathbf{r}_{\mathrm{b}},\mathbf{X} \right) \mathbf{h}_{\mathrm{b}}^{\mathrm{H}}\left( \mathbf{r}_{\mathrm{b}},\mathbf{X} \right) \notag \\
    &\qquad +\sigma _{\rm b}^{2}\mathbf{H}_{\mathrm{w}}^{\mathrm{H}}\left( \hat{\mathbf{v}}_{\mathrm{w}},\hat{\mathbf{r}}_{\mathrm{w}},\mathbf{X} \right) \mathbf{H}_{\mathrm{w}}^{}\left( \hat{\mathbf{v}}_{\mathrm{w}},\hat{\mathbf{r}}_{\mathrm{w}},\mathbf{X} \right),  \\
    c\left( \mathbf{X} \right) &\triangleq \left| \mathbf{h}_{\mathrm{b}}^{\mathrm{H}}\left( \mathbf{r}_{\mathrm{b}},\mathbf{X} \right) \tilde{\mathbf{w}} \right|^2.
\end{align}
The form of the objective function \eqref{objective_sensing_w_sub1} aligns with the definition of generalized Rayleigh quotient.
However, note that, although $\mathbf{B}\left( \mathbf{X} \right) \succeq 0 $ holds for any $\tilde{\mathbf{q}}$, the Rayleigh quotient requires $\mathbf{B}\left( \mathbf{X} \right) \succ 0 $ to guarantee the solvability of the problem.
Therefore, we need to discuss the case that $\tilde{\mathbf{q}}^{\mathrm{H}}\mathbf{B}\left( \mathbf{X} \right) \tilde{\mathbf{q}} = 0$.
In this case, we have
\begin{align}
    &\Gamma _{\mathrm{sen}}\tilde{\mathbf{q}}^{\mathrm{H}}\mathbf{h}_{\mathrm{b}}^{}\left( \mathbf{r}_{\mathrm{b}},\mathbf{X} \right) \mathbf{h}_{\mathrm{b}}^{\mathrm{H}}\left( \mathbf{r}_{\mathrm{b}},\mathbf{X} \right) \tilde{\mathbf{q}} \notag \\
    &\qquad \qquad +\sigma _{\rm b}^{2}\tilde{\mathbf{q}}^{\mathrm{H}}\mathbf{H}_{\mathrm{w}}^{\mathrm{H}}\left( \hat{\mathbf{v}}_{\mathrm{w}},\hat{\mathbf{r}}_{\mathrm{w}},\mathbf{X} \right) \mathbf{H}_{\mathrm{w}}^{}\left( \hat{\mathbf{v}}_{\mathrm{w}},\hat{\mathbf{r}}_{\mathrm{w}},\mathbf{X} \right) \tilde{\mathbf{q}}=0. \notag
\end{align}
The above equation holds when 
\begin{align}
    &\tilde{\mathbf{q}}^{\mathrm{H}}\mathbf{h}_{\mathrm{b}}^{}\left( \mathbf{r}_{\mathrm{b}},\mathbf{X} \right) \mathbf{h}_{\mathrm{b}}^{\mathrm{H}}\left( \mathbf{r}_{\mathrm{b}},\mathbf{X} \right) \tilde{\mathbf{q}}=0, \notag \\
    &\tilde{\mathbf{q}}^{\mathrm{H}}\mathbf{H}_{\mathrm{w}}^{\mathrm{H}}\left( \hat{\mathbf{v}}_{\mathrm{w}},\hat{\mathbf{r}}_{\mathrm{w}},\mathbf{X} \right) \mathbf{H}_{\mathrm{w}}^{}\left( \hat{\mathbf{v}}_{\mathrm{w}},\hat{\mathbf{r}}_{\mathrm{w}},\mathbf{X} \right) \tilde{\mathbf{q}}=0, \notag
\end{align}
which indicates that $\tilde{\mathbf{q}}$ simultaneously falls into the orthogonal space spanned by $\mathbf{h}_{\mathrm{b}}^{}\left( \mathbf{r}_{\mathrm{b}},\mathbf{X} \right) \mathbf{h}_{\mathrm{b}}^{\mathrm{H}}\left( \mathbf{r}_{\mathrm{b}},\mathbf{X} \right) $ and $\mathbf{H}_{\mathrm{w}}^{\mathrm{H}}\left( \hat{\mathbf{v}}_{\mathrm{w}},\hat{\mathbf{r}}_{\mathrm{w}},\mathbf{X} \right) \mathbf{H}_{\mathrm{w}}^{}\left( \hat{\mathbf{v}}_{\mathrm{w}},\hat{\mathbf{r}}_{\mathrm{w}},\mathbf{X} \right) $.
However, in this case, the numerator of the objective function \eqref{objective_sensing_w_sub1} can be computed as 
\begin{align}
    \tilde{\mathbf{q}}^{\mathrm{H}}\mathbf{A}\left( \mathbf{X} \right) \tilde{\mathbf{q}}& \notag =\tilde{\mathbf{q}}^{\mathrm{H}}\mathbf{H}_{\mathrm{w}}^{\mathrm{H}}\left( \hat{\mathbf{v}}_{\mathrm{w}},\hat{\mathbf{r}}_{\mathrm{w}},\mathbf{X} \right) \mathbf{H}_{\mathrm{w}}^{}\left( \hat{\mathbf{v}}_{\mathrm{w}},\hat{\mathbf{r}}_{\mathrm{w}},\mathbf{X} \right) \tilde{\mathbf{q}} \notag \\&\qquad \qquad \qquad -\Gamma _{\mathrm{sen}}\tilde{\mathbf{q}}^{\mathrm{H}} \tilde{\mathbf{q}} = -\Gamma _{\mathrm{sen}} < 0, \notag
\end{align}
which is undesirable for throughput maximization.
Furthermore, the following lemma rigorously proves that the entire $\tilde{\mathbf{q}}$ must lies in the parallel subspace.
\begin{lemma} \label{lemma:1}
The direction of the optimal solution $\tilde{\mathbf{q}}^\star$ to \eqref{objective_sensing_w_sub1} must be located in the subspace spanned by $\mathbf{H}_{\mathrm{w}}^{}\left( \hat{\mathbf{v}}_{\mathrm{w}},\hat{\mathbf{r}}_{\mathrm{w}},\mathbf{X} \right) $ and $\mathbf{h}_{\mathrm{b}}^{}\left( \mathbf{r}_{\mathrm{b}},\mathbf{X} \right) $. 
\begin{IEEEproof}
    Please refer to Appendix \ref{appendix:subspace}.
\end{IEEEproof}
\end{lemma}
Therefore, the optimal solution for $\tilde{\mathbf{q}}$ cannot be located in the orthogonal space of $\mathbf{B}\left( \mathbf{X} \right) $.
Accordingly, eigenvalue decomposition (EVD) can be applied on $\mathbf{B}\left( \mathbf{X} \right) $ to extract its positive subspace.
In particular, $\mathbf{B}\left( \mathbf{X} \right) $ can be decomposed as $\mathbf{B}\left( \mathbf{X} \right) =\mathbf{U\Lambda U}^{\mathrm{H}}
$ by EVD.
Then, we extract the non-zero eigenvalues in $\mathbf{\Lambda }$ and their corresponding eigenvectors in $\mathbf{U}_+$ to form the positive definite subspace, which can be expressed as $\mathbf{B}_+\left( \mathbf{X} \right) =\mathbf{U}_+\mathbf{\Lambda }_+\mathbf{U}_{+}^{\mathrm{H}}$, where $\mathbf{B}_+\left( \mathbf{X} \right) \succ 0$.
Therefore, letting $\tilde{\mathbf{q}}\triangleq \mathbf{U}_{+}^{\mathrm{H}}\mathbf{\Lambda }_{+}^{-1/2}\mathbf{q}^{\prime}
$, the original Rayleigh quotient can be recast as 
\begin{subequations}\label{problem:rayleigh_quotient}
\begin{align}
    \max_{\mathbf{q}^{\prime}} \quad c\left( \mathbf{X} \right) \frac{\left( \mathbf{q}^{\prime} \right) ^{\mathrm{H}}\mathbf{C}\left( \mathbf{X} \right) \mathbf{q}^{\prime}}{\left( \mathbf{q}^{\prime} \right) ^{\mathrm{H}}\mathbf{q}^{\prime}}, \label{objective_sensing_w_sub2}
\end{align}
\end{subequations}
where matrix $\mathbf{C}\left( \mathbf{X} \right)$ is defined as 
\begin{align}
    \mathbf{C}\left( \mathbf{X} \right) \triangleq \mathbf{\Lambda }_{+}^{-1/2}\mathbf{U}_{+}^{\mathrm{H}}\mathbf{A}\left( \mathbf{X} \right) \mathbf{U}_{+}^{\mathrm{H}}\mathbf{\Lambda }_{+}^{-1/2}. \label{eq:solution_to_q}
\end{align}

\subsubsection{Subproblem with respect to $\mathcal{X}$}
Once the beamforming vector $\mathbf{w}_t$ and $\mathbf{q}_t$ for $\forall t$ are obtained via the methods presented in sub-sections \ref{sect:subproblem_wrt_W} and \ref{sect:subproblem_to_Q}, the remaining part of the original problem \eqref{problem:original} is the optimization of the antenna position matrix set $\mathcal{X}$, which can be specified by
\begin{subequations}
\label{subproblem:X}
\begin{align}
  \max_{\mathcal{X}}\quad
    & R(\mathcal{X}) \label{objective_X}\\[6pt]
  \text{s.t.} \quad &\eqref{constraint:position_1}\text{~and~}\eqref{constraint:position_2}
\end{align}
\end{subequations}
In dynamic systems, the real-time PA position optimization is challenging.
To address this issue, we consider a sub-array layout, where the positions of the initial PAs on each waveguide are optimized, while keeping the inter-spacings between PAs fixed but larger than the minimal allowed distance $\Delta_{\min}$.
The reason for adopting this layout is that a dynamic system is considered in this work, where real-time optimization is necessary.
By fixing the inter-spacings between PAs, the optimization complexity decreases from quadratic to linear, thereby striking a balance between performance and complexity.
Hence, the subproblem \eqref{subproblem:X} can be recast as 
\begin{subequations}
\label{subproblem:X-sub1}
\begin{align}
  \max_{\mathcal{X}_{{\mathrm{init}}}}\quad
    & R(\mathcal{X}_{{\mathrm{init}}}) \label{objective_x_init}\\
  \text{s.t.} \quad &\left[ \mathbf{x}_{\mathrm{init},t} \right] _n \in \left[ 0,L_{\max}-\left( M_{\mathrm{t}}-1 \right) \Delta _x \right], \forall t, n, \label{constraint:x_init}
\end{align}
\end{subequations}
where $\mathbf{x}_{{\mathrm{init}, t}} \triangleq [x_{1,1,t}, x_{2,1,t}, ..., x_{N_{\rm t},1,t}]^{\rm T} \in \mathbb{R}^{N_{\rm t} \times 1}$ is the $x$-coordinate vector of the initial PAs on all waveguides, and $\Delta _x \ge \Delta_{\min}$ represents the uniform inter-spacing between adjacent PAs on the same waveguide.

Since Willie's trajectory can be modeled as a time-related process, the temporal information between CPIs can be leveraged.
In other words, the optimization of the current PA position matrix $\mathbf{X}_{t}$ at the $t$-th CPI is relevant to that in the former CPI, i.e., $\mathbf{X}_{t-1}$.
To effectively capture this relationship, we employ DRL in the optimization of $\mathcal{X}$, whose core idea is to learn the hidden pattern or rules underlying the evolution of an environment through exploitation and exploration.
Specifically, we adopt the soft actor-critic (SAC) approach to solve problem \eqref{subproblem:X-sub1}.
In contrast to conventional DRL approaches, such as deep Q-learning (DQN), deep deterministic policy gradient (DDPG), and proximal policy optimization (PPO), etc, the SAC approach introduces an entropy regularization term in the optimization function to control exploitation and exploration \cite{jiang2024fairness, haarnoja2019soft}. 
The entropy term measures the randomness of the action distributions.
Thus, weighting the entropy term more heavily encourages a wider variety of actions, thereby emphasizing exploration.
Furthermore, by treating the weight factor of the entropy term as a trainable parameter, the performance of SAC will be more robust to the choice of hyperparameters.
In what follows, we present how to tailor the canonical SAC for solving problem \eqref{subproblem:X}, detailing \emph{1) Settings of State, Action, and Reward}, and \emph{2) Framework of SAC}.

\textbf{1)~Settings of State, Action, and Reward:}
The setting of the state should let the agent, i.e., Alice, comprehensively perceive the environment and how it evolves, which is related to the movement of Willie.
Therefore, as the channels are LoS and parameterized by the positions of Willie, Bob, and PAs amounted on Alice's PASS, we define a state at the $t$-th CPI as $\mathbf{s}_{t} = [\mathbf{r}_{\rm b}^{\mathrm{T}}, \hat{\mathbf{r}}_{\mathrm{w}, t}^{\mathrm{T}}, \mathbf{x}_{{\mathrm{init}}, t}^{\rm T}] \in \mathbb{R}^{(2+2+N_{\rm t}) \times 1}$, where the $z$-coordinates of $\mathbf{r}_{\rm b}$ and $\hat{\mathbf{r}}_{\mathrm{w}, t}$ are omitted.
Then, the definition of actions is straightforward and specified by $\mathbf{a}_t = \mathbf{x}_{{\mathrm{init}},t}$, i.e., the optimization variable in problem \eqref{subproblem:X-sub1}.
It is noteworthy that the action will be clipped to the range $[0, L_{\max}-\left( M_{\mathrm{t}}-1 \right) \Delta _x]$ by the activation function to meet the constraint \eqref{constraint:x_init}. 
Finally, the reward function is set to the instantaneous covert rate given by $r(\mathbf{s}_t, \mathbf{a}_t) = \log _2\left( 1+\gamma _t\left( \mathbf{w}_t,\mathbf{q}_t,\mathbf{x}_{\mathrm{init},t} \right) \right)$. 

\textbf{2)~Framework of SAC:}
The objective of SAC can be expressed as
\begin{align}
    \pi ^{\star}=\mathrm{arg}\max _{\pi}\mathbb{E} _{\pi}\left\{ \sum\nolimits_t^{}{r(\mathbf{s}_t,\mathbf{a}_t)+\varphi \mathcal{H} \left( \pi \left( \cdot \mid \mathbf{s}_t \right) \right)} \right\}, \label{eq:sac_obj}
\end{align}
where $\mathcal{H} \left( \pi \left( \cdot \mid \mathbf{s}_t \right) \right) =-\log \pi \left( \cdot \mid \mathbf{s}_t \right)$ denotes the entropy of policy $\pi$ at state $\mathbf{s}_t$, and $\varphi$ is the weight factor for the entropy regularization term.
In general, the structure of SAC contains two modules: 1) the actor module, and 2) the critic module.
The actor module can be modeled as a function parameterized by $\boldsymbol{\theta}$.
The input of this module is a state $\mathbf{s}_t$, and the output of this module is the action distribution over the whole action space.
The specified action will be sampled from the action distribution, i.e., $\mathbf{a}_t \sim \pi \left( \cdot \mid \mathbf{s}_t;\boldsymbol{\theta } \right)$.
However, the direct sampling operation from the action distribution will prevent the propagation of the gradient.
To address the problem, a reparameterization trick is employed here.
In particular, the actor module will output the distribution's mean $\boldsymbol{\mu}_\theta(\mathbf{s}_t)$ and standard deviation $\boldsymbol\sigma_\theta(\mathbf{s}_t)$.
Then, the real action can be generated by 
\begin{align}
  \tilde{\mathbf{a}}_t
  &= \boldsymbol{\mu}_\theta(\mathbf{s}_t)
     + \boldsymbol\sigma_\theta(\mathbf{s}_t)\odot \epsilon_t,
  \quad \epsilon_t \sim \mathcal{N}(\mathbf{0}, \mathbf{I}_{N_{\rm t}}). \label{eq:reparam}
\end{align} 
To train the actor module, the loss function is given by
\begin{align}
    \mathcal{L} _{\pi}\left( \boldsymbol{\theta } \right) =\mathbb{E} _{\mathcal{D} \sim \mathcal{Z}}\left\{ \varphi \log \pi \left( \tilde{\mathbf{a}}_i\mid \mathbf{s}_i \right) -\underset{j=1,2}{\min}Q_j\left( \mathbf{s}_i,\tilde{\mathbf{a}}_i;\boldsymbol{w}_j \right) \right\}, \label{eq:actor_loss}
\end{align}
where $\mathcal{Z} \triangleq \left\{\left( \mathbf{s}_i,\tilde{\mathbf{a}}_i,r\left( \mathbf{s}_i,\tilde{\mathbf{a}}_i \right) ,\mathbf{s}_{i+1} \right)_i \right\} $ denotes the replay buffer, which stores each transition tuple—state, action, received reward, and next state—experienced by the agent; and $Q_j\left( \mathbf{s}_i,\tilde{\mathbf{a}}_i;\boldsymbol{w}_j \right)$ denotes the output of the $j$-th critic module, which will be elaborated later.
Here, $\mathcal{D} \sim \mathcal{Z}$ indicates that a mini-batch $\mathcal{D} $ with cardinality of $|\mathcal{D}|$ is randomly sampled from $\mathcal{Z}$ for training.
SAC, as an off-policy DRL approach, can learn from its gained experience in $\mathcal{Z}$ to improve the data efficiency, which is described as experience replay.
From \eqref{eq:actor_loss}, we notice that two critic networks indexed by $j=1,2$ are utilized.
The reason lies in the fact that the Q-value overestimation problem can be overcome by always taking the smaller value among the outputs of the two critic modules.
Then, upon the loss function in \eqref{eq:actor_loss}, the update rule for the trainable parameter $\boldsymbol{\theta}$ is given by
\begin{align}
    \boldsymbol{\theta }\gets \boldsymbol{\theta }-w_{\theta}\nabla \mathcal{L} _{\pi}\left( \boldsymbol{\theta } \right), \label{eq:update_actor}
\end{align}
wherein $w_{\theta} > 0$ denotes the learning rate.

Next, we introduce the critic model, which is used to evaluate an action-state pair concerning its long-term reward.
Given the input is an action-state pair, the critic module can be regarded as a mapping function parameterized by $\boldsymbol{w}$ and expressed as $Q_j\left( \mathbf{s}_i,\tilde{\mathbf{a}}_i;\boldsymbol{w}_j \right)$ with $j=\{1,2\}$ indicating a dual critic structure.
To stabilize the training process, two target critic modules are utilized, denoted by $Q_{j}^{\left( \mathrm{tar} \right)}( \mathbf{s}_i,\tilde{\mathbf{a}}_i;\boldsymbol{w}_{j}^{( \mathrm{tar} )} ) $ with $j \in \{1, 2\}$.
At each update, the trainable parameters of the target critic modules will be softly synchronized with that of the training critic modules via $\boldsymbol{w}_{j}^{\left( \mathrm{tar} \right)}\gets \tau \boldsymbol{w}_{j}+(1-\tau)\boldsymbol{w}_{j}^{\left( \mathrm{tar} \right)}$, where $\tau \in (0, 1)$ denotes the weight factor.
The loss function of the critic model is specified by
\begin{align}
    \mathcal{L} _Q\left( \boldsymbol{w}_j \right) =\mathbb{E} _{\mathcal{D} \sim \mathcal{Z}}\left\{ \frac{1}{2}\left( Q_j\left( \mathbf{s}_i,\tilde{\mathbf{a}}_i;\boldsymbol{w}_j \right) -y_{j}^{\left( \mathrm{tar} \right)} \right)^2 \right\},
\end{align}
where 
\begin{align}
    y_{j}^{\left( \mathrm{tar} \right)}&=r\left( \mathbf{s}_i,\tilde{\mathbf{a}}_i \right) +\gamma \min_{j=\left\{ 1,2 \right\}} Q_{j}^{\left( \mathrm{tar} \right)}(\mathbf{s}_{i+1},\tilde{\mathbf{a}}_{i+1};\boldsymbol{w}_{j}^{(\mathrm{tar)}}) \notag \\
    &\qquad \qquad \qquad \qquad \qquad -\varphi \log \pi \left( \tilde{\mathbf{a}}_{i+1}\mid \mathbf{s}_{i+1} \right), \label{eq:loss_func_critic}
\end{align}
and $\gamma$ denotes the discount factor used for the evaluation of the importance of the future rewards.
Based on the loss function in \eqref{eq:loss_func_critic} the update rule for $\boldsymbol{w}$ can be expressed as
\begin{align}
    \boldsymbol{w}_{j}^{}\gets \boldsymbol{w}_{j}^{}-w_w\nabla \mathcal{L} _Q\left( \boldsymbol{w}_{j}^{} \right),\label{eq:update_critic}
\end{align}
where $w_w>0$ denotes the learning rate for the critic module.

Finally, the loss function of the weight factor $\varphi$ in \eqref{eq:sac_obj} can be expressed as
\begin{align}
    \mathcal{L} _{\varphi}\left( \varphi \right) =\mathbb{E} _{\mathcal{D} \sim \mathcal{Z}}\left\{ -\varphi \log \pi \left( \tilde{\mathbf{a}}_i\mid \mathbf{s}_i \right) -\varphi \mathcal{H} _0 \right\},
\end{align}
where $\mathcal{H}_0$ denotes the target entropy.
Finally, the update rule for this term is given by
\begin{align}
    \varphi \gets \varphi -w_{\varphi}\nabla \mathcal{L} _{\varphi}\left( \varphi \right),\label{eq:update_weight}
\end{align}
where $w_{\varphi} >0$ denotes the learning rate.

The overall algorithm is summarized in \textbf{Algorithm \ref{alg:overall}}.
\begin{algorithm}[t!]
	\SetAlgoLined
	\small
	\caption{DRL-Based Solution to Problem \eqref{problem:original}}
	\label{alg:overall}
	\KwIn{
		Precise position of Bob $\mathbf{r}_{\mathrm{b}}$; 
		Initial mobility status of Willie given by $\boldsymbol{\xi}_0$; 
		Total power budget $P_{\max}$ and sensing threshold $\Gamma_{\rm sen}$;
            Noise power $\sigma^2_{\mathrm{w}}$, $\sigma^2_{\mathrm{b}}$, and $\sigma^2_{\mathrm{s}}$;
		Initialized SAC hyper parameters $w_\theta$, $w_w$, and $w_{\varphi}$, hidden dimensional, target entropy $\mathcal{H}_0$, replay buffer size $| \mathcal{Z} |$, and $| \mathcal{D} |$.
	}
	\DontPrintSemicolon
	
	\textbf{Initialization}: 
	Set iteration index $t \leftarrow 0$;
	Randomly initialize the positions of the weights of actor and critic modules $\boldsymbol{\theta}$, $\boldsymbol{w}_1$, $\boldsymbol{w}_2$, $\boldsymbol{w}_1^{\mathrm{(tar)}}$, and $\boldsymbol{w}_2^{\mathrm{(tar)}}$\;
        Establish the NN architecture $\pi \left( \cdot \mid \mathbf{s}_t;\boldsymbol{\theta } \right) $, $Q_1( \mathbf{s}_t,\tilde{\mathbf{a}}_t;\boldsymbol{w}_1)$ and $Q_2( \mathbf{s}_t,\tilde{\mathbf{a}}_t;\boldsymbol{w}_2 ) $.
        The target critic model $Q_1^{\mathrm{(tar)}}( \mathbf{s}_t,\tilde{\mathbf{a}}_t;\boldsymbol{w}_1^{(\mathrm{tar})} )$ and $Q_2^{\mathrm{(tar)}}( \mathbf{s}_t,\tilde{\mathbf{a}}_t;\boldsymbol{w}_2^{(\mathrm{tar})}) $\;
        Set the minimal memory load allowing for training as $Z_{\min}$
	Randomly initialize the PAs' position for the first CPI\;
	
	\For{$t=1$ \KwTo\ $T$}{
        \tcp{Prior Prediction Via EKF:}
	Obtain $\hat{\boldsymbol{\xi}}_t$ via the kinematic model \eqref{eq:prior_mobility}\;
        Obtain $\mathbf{P}_{t \mid t-1}$ via the covariance estimation rule \eqref{eq:coverianxce_prior} \;
        Compute Willie's CSI via signal model \eqref{eq:willie_cis} \;

        \tcp{DRL Solution Via EKF Estimations:}
        \tcp{1)~Antenna Position Optimization: }
        Construct current state vector $\mathbf{s}_t$ \; 
        Feed the actor module $\pi \left( \cdot \mid \mathbf{s}_t;\boldsymbol{\theta } \right)$ with $\mathbf{s}_t$\;
        Obtain the action vector via the reparameterization trick in \eqref{eq:reparam}\;
        \tcp{2)~Beamforming and AN Design:}
        Construct the unit-power beamformer $\tilde{\mathbf{w}}_t$ via \eqref{eq:solution_to_w} \;
        Obtain the AN vector $\mathbf{q}_t$ by solving \eqref{problem:rayleigh_quotient}\; 
        Obtain beamforming vector ${\mathbf{w}}_t$ by allocating the rest power to $\tilde{\mathbf{w}}_t$ \; 
        Obtain the reward $r_t$ by executing the action $\tilde{\mathbf{a}}_t$ and transmitting using $\mathbf{c}_t=\mathbf{w}_t s_t + \mathbf{q}_t$\;
        Enter the next state $\mathbf{s}_{t+1}$ \;
        \tcp{3) Training Step of DRL:}
        Place memory $\{ \mathbf{s}_t, \tilde{\mathbf{a}}_t, r_t, \mathbf{s}_{t+1} \}$ into the memory pool $\mathcal{Z}$ \;
        \If{$|\mathcal{Z}| \ge Z_{\min}$}{
            Randomly sample a mini-batch of experience $\mathcal{D}$ of a size $|\mathcal{D}|$ \;
            Update the critic module according to \eqref{eq:update_critic}\;
            Update the actor module according to \eqref{eq:update_actor} \;
            Update the regularization term according to \eqref{eq:update_weight}\;
            Synchronize the target and training critic modules \;
         }   
        \tcp{Posterior Update of EKF: }
        Receive the real echo signal $\mathbf{y}_{\mathrm{w},t}$\;
        Compute the residual covariance matrix via \eqref{eq:S}\;
        Compute the Kalman gain via \eqref{eq:kalman_gain}\;
        Update the mobility status via \eqref{eq:update_mobility} \;
        Update the posterior covariance matrix via \eqref{eq:covariance_update}\;
        Enter next CPI via $t\leftarrow t+1$ \;
	}
\end{algorithm}

\section{Numerical Results} \label{sect:results}
In this section, the simulation results are presented to evaluate the performance of the proposed approach.
The following simulation setups are utilized throughout this section unless stated otherwise.

\subsection{Parameter Settings and Benchmarks}
For the basic parameters, the carrier frequency is set to 15 GHz \cite{wang2025modeling} with the effective refraction index in waveguides and LCXs being set to $1.4$ \cite{wang2025modeling} and $1.1$ \cite{ji2024theoretical}, respectively.
The noise power density is set to $-174~\rm{dBm/Hz}$ with $10~\rm{kHz}$ narrowband for transmission and reception.
On the transmitter side, the number of waveguides and the number of PAs on each waveguide are set to $N_{\rm t} = 3$ and $M_{\rm t}=4$, respectively.
On the sensing side, the number of slots on LCXs is distributed with a uniform spacing of $1~\mathrm{m}$, and the number of LCX for echo reception is set to $N_{\rm r} = 3$.
The heights of PASS and LCX are uniformly set to $H=3~\mathrm{m}$.
The spacing between two adjacent waveguides and that between two adjacent LCXs are uniformly fixed as $5~{\mathrm{m}}$.
The distance between the waveguide and the neighborhood LCX is set to $0.5~\mathrm{m}$.
For the sensing-related parameters, the duration of each CPI is specified by $\Delta T = 0.1~\mathrm{ms}$, and the reflection coefficient is set to a constant number \cite{yuan2021bayesian}.
Without loss of generality, we fix $\beta=1$.
The transmit power is set to $30~\mathrm{dBm}$ with the sensing threshold being specified to $-50~{\mathrm{dBm}}$.
Finally, the hyperparameters for the SAC approach are summarized in Table \ref{tab:sac_hyperparams}.

\begin{table}[t!]
  \centering
  \caption{Hyper-parameters for the SAC agent}\label{tab:sac_hyperparams}
  \footnotesize
  \renewcommand{\arraystretch}{1.1}
  \begin{tabularx}{\linewidth}{@{}l c l c@{}}
    \toprule
    \textbf{Parameter} & \textbf{Value} & \textbf{Parameter} & \textbf{Value} \\ \midrule
    State dim            & 7               & Hidden dim            & 128          \\
    Action dim           & 3               & Action bound          & 1            \\
    Actor LR             & $3\times10^{-4}$& Critic LR             & $3\times10^{-3}$ \\
    $\alpha$ LR          & $3\times10^{-4}$& Target entropy        & $-3$         \\
    Discount factor $\tau$ & 0.005           & Discount factor $\gamma$ & 0.99     \\
    Batch size           & 64              & Buffer size           & 1\,000       \\
    Minimal size         & 100             &                       &              \\ \bottomrule
  \end{tabularx}
\end{table}
The following benchmark algorithms are used for validating the effectiveness of the proposed method.
\begin{itemize}
\item \textbf{PASS, One-Dimensional (1D) Search}:  
This baseline replaces the SAC approach with an exhaustive 1D search for the PA position.  
Because an exhaustive search is computationally intensive, we set the search step to $L_{\max}/10$.  
With this resolution, the running time of the 1D search is comparable to that of the SAC approach, ensuring a fair comparison.

\item \textbf{PASS, Greedy}:  
Given Willie’s true trajectory, this heuristic baseline places the initial PAs on the waveguides at Bob's $x$-coordinate, to greedily enlarge the throughput at Bob, while the remaining PAs are then spaced at half-wavelength intervals.  
This simple rule of design provides a low-complexity, myopic heuristic approach for PA position optimization.

\item \textbf{MIMO, Perfect CSI}:  
The conventional full-digital MIMO array is equipped with $N_{\rm t}$ half-wavelength fixed antenna elements and centered at $[0.0~\mathrm{m},\,0.0~\mathrm{m},\,3.0~\mathrm{m}]^{\mathrm{T}}$.  
When the aperture size of the receiver array is small, a standalone MIMO system struggles to track Willie with nearly planar waves. 
To isolate the sensing contribution, we therefore fed this MIMO benchmark with the ground-truth trajectory of Willie, indicating that perfect CSI is available at Alice.
\end{itemize}

\subsection{Tracking Performance of EKF Method}
\begin{figure}[t!]
    \centering
    \includegraphics[width=0.80\linewidth]{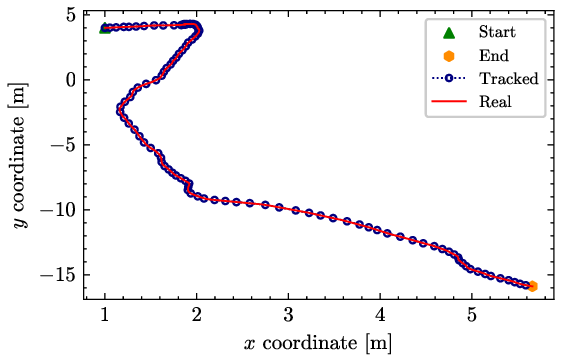}
    \caption{Illustration of tracking results on Willie's trajectory.}
    \label{fig:trajectory_tracking}
    \vspace{-10pt}
\end{figure}
\begin{figure}[htbp]
  \centering
  \subfloat[$x$-direction velocity tracking\label{fig:x_velo_tracking}]{
      \includegraphics[width=0.85\linewidth]{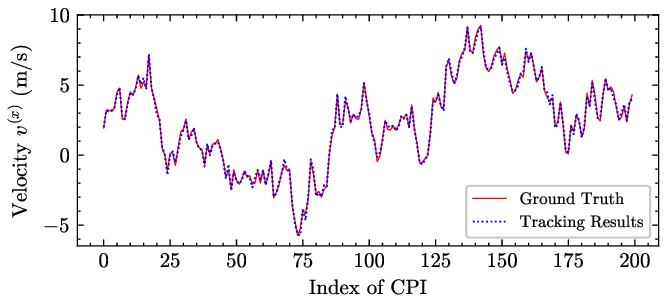}
  }
  \hfill 

  \subfloat[$y$-direction velocity tracking\label{fig:y_velo_tracking}]{
      \includegraphics[width=0.85\linewidth]{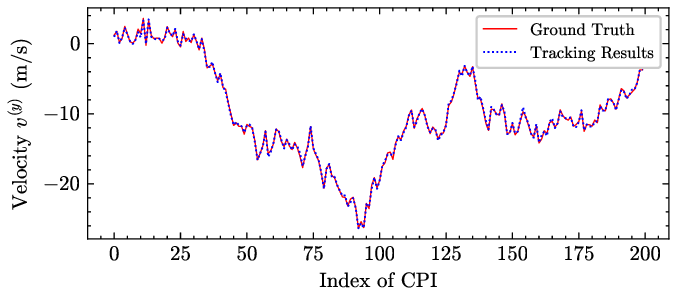}
  }
  \caption{Velocity-tracking performance in the $x$ and $y$ directions.}
  \label{fig:velo_tracking}
  \vspace{-10pt}
\end{figure}
\begin{figure}[t!]
    \centering    \includegraphics[width=0.7\linewidth]{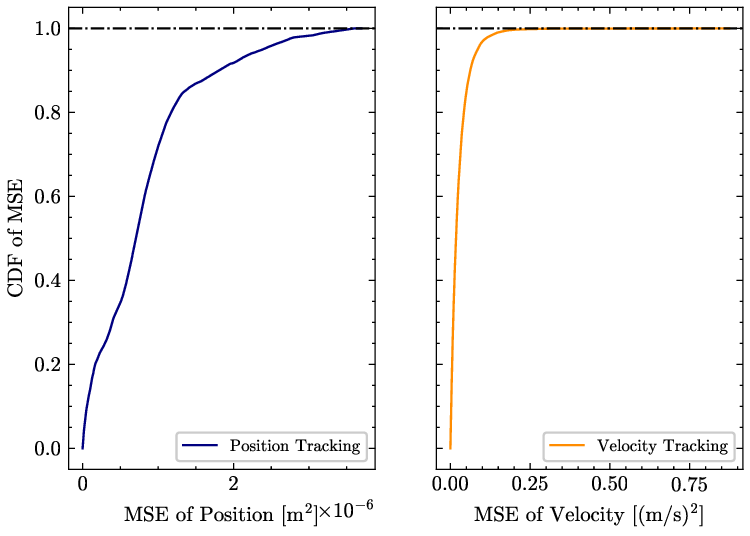}
    \caption{Illustration of the CDF of the MSE error in position and velocity tracking.}
    \label{fig:cdf-mse}
    \vspace{-10pt}
\end{figure}

The initial position of Willie is set to $[1~\mathrm{m}, 4~\mathrm{m}, 0~\mathrm{m}]^T$ and the initial velocities are set to $v_x = 2~\mathrm{m/s}$ and $v_y = 1~\mathrm{m/s}$.
The trajectory is generated randomly with velocity variance $\sigma_{v_x}^2 = 0.01$ and $\sigma_{v_y}^2 = 0.02$.
Under this setup, Fig. \ref{fig:trajectory_tracking} shows that the proposed EKF method can accurately track the position of Willie, even though a ``U"-shape turn exists.
It is essential to note that the extended near-field region resulting from the large aperture size of PASS enables full-dimensional tracking by exploiting spherical waves.
In addition to the trajectory tracking, the velocity tracking results are shown in Fig. \ref{fig:velo_tracking}, which also demonstrates the high-fidelity sensing performance achieved by the EKF method.

To gain a comprehensive knowledge of errors in the mobility status tracking process, Fig. \ref{fig:cdf-mse} shows the empirical cumulative density function (CDF) of the mean square error (MSE) in tracking results compared to the ground truth.
From this figure, it can be observed that the position tracking MSE is at the $10^{-6}$ level, while the velocity tracking MSE is at the $10^{-1}$ scale.
This is because, before the posterior update step in EKF, the velocity prediction is conducted purely on the inaccurate kinematic model that excludes the unobservable velocity variances.
However, due to the short duration of each CPI, the error in velocity predictions has little effect on the overall results.
Moreover, compared to ELAA-enabled near-field sensing, which often relies on hundreds or even thousands of half-wavelength antenna elements, the computational complexity of EKF for PASS is moderate because a broad near-field region can be formed with only a handful of sparsely distributed PAs.

\subsection{Covert Rate Performance}
\begin{figure}[t!]
    \centering
    \includegraphics[width=0.85\linewidth]{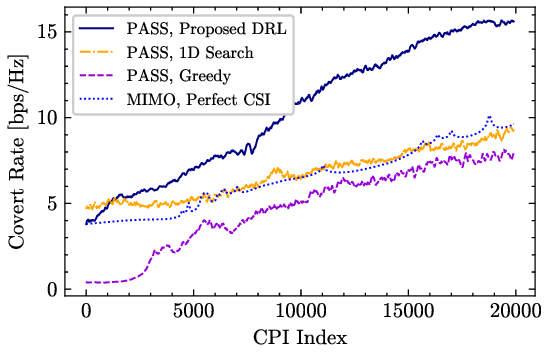}
    \caption{Illustration of the covert rate tracking over the trajectory shown in Fig. \ref{fig:trajectory_tracking}.}
    \label{fig:covert_rate_one_trajectory}
    \vspace{-10pt}
\end{figure}
\begin{figure}[t!]
    \centering
    \includegraphics[width=0.85\linewidth]{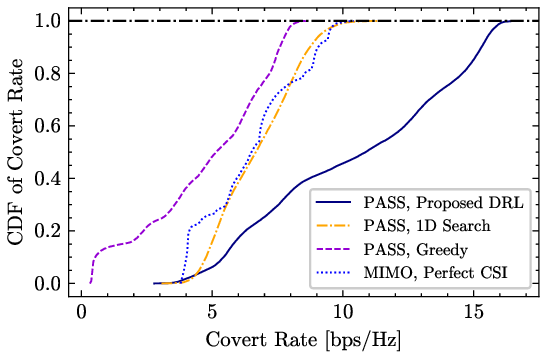}
    \caption{Illustration of the empirical CDF of the achieved covert rate over the trajectory shown in Fig. \ref{fig:trajectory_tracking}.}
    \label{fig:cdf_covert_rate}
    \vspace{-10pt}
\end{figure}
\begin{figure}[t!]
    \centering
    \includegraphics[width=0.85\linewidth]{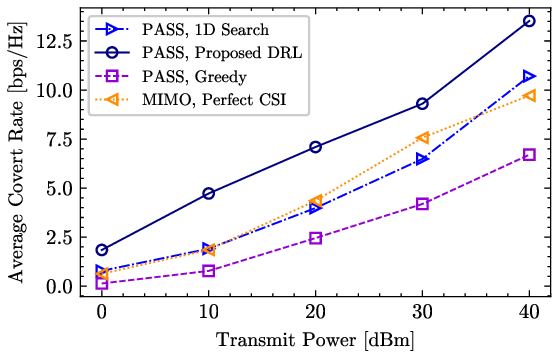}
    \caption{Illustration of the averaged covert rate tracking over the trajectory shown in Fig. \ref{fig:trajectory_tracking} under different transmit power levels. }
    \label{fig:covert_rate_diff_power}
    \vspace{-10pt}
\end{figure}

As sensing results are verified, the CSI at Willie can be obtained correspondingly. 
Building on this, the covert rate performance is investigated in this subsection.
For the parameter setups, Bob's position is fixed at $\mathbf{r}_{\mathrm{b}}=[3~\mathrm{m}, 5~\mathrm{m}, 0~\mathrm{m}]^{\rm T}$.
To circumvent incidental results, all results are obtained through $20$ Monte Carlo simulations.
Additionally, a moving average with a window size of 100 is employed to capture hidden patterns through CPIs, a technique often used in DRL simulations.

Fig. \ref{fig:covert_rate_one_trajectory} illustrates the covert rate tracking results when the trajectory of Willie follows that shown in Fig. \ref{fig:trajectory_tracking}.
It can be observed that as the CPI index increases, the proposed DRL-based algorithm gradually learns to tune the position of PAs based on the evolution of Willie's mobility status, ultimately leading to significant outperformance over the other benchmarks.

Regarding the benchmarks, we can observe that the covert rate achieved by all baseline approaches increases over time.
This is because, as Willie moves away from Bob, a higher percentage of power can be allocated to Bob, thus leading to an increase in the covert rate.
Compared to the ``PASS, Greedy" approach, the superior performance of the DRL method highlights the important role that learning plays in optimization over a temporal sequence, underscoring the necessity of dedicated, practical algorithms for PASS.
The ``PASS, 1D Search" algorithm fails to capture the temporal information between CPIs and treats CPIs independently.
In this case, the superiority of learning is proved by this comparison.
The performance of the ``PASS, 1D Search" baseline can be enhanced by increasing the searching resolution by dividing the searching positions of PAs into finer grids.
However, the correspondingly increased computational complexity will prevent its practical deployments, especially for this dynamic network. 
Finally, the performance gain by adopting PASS is verified by comparing it with the conventional ``MIMO, Perfect CSI" benchmark.
To gain a close look at the achieved covert rate alongside the trajectory, the empirical CDF of the covert rate is demonstrated in Fig. \ref{fig:cdf_covert_rate}.
Finally, we investigate the performance of the proposed algorithm as a function of transmit power.
As shown in Fig. \ref{fig:covert_rate_diff_power}, it can be observed that the proposed DRL method, as well as the EKF tracking module, are robust to varying transmit power.

\vspace{-1em}
\section{Conclusion}\label{sect:conclusion}
The sensing-assisted covert communication system empowered by PASS was investigated in this work, where the adversary user's CSI was unknown to the transmitter.
To tackle this problem, the sensing functionality using EKF was leveraged to track Willie and obtain its CSI, thus facilitating the covert transmission to the legitimate user.
Based on the sensing results, the covert communication problem was formulated and divided into three sub-problems: beamforming, AN design, and PA position optimization.
Using the subspace method, the beamforming and AN design problem was optimally solved.
Building on the above, the SAC approach was employed to solve the PA position optimization sub-problem to capture the temporal correlation between CPIs.
Numerical results were presented to show the effectiveness of the proposed solution while validating the superiority of PASS over conventional MIMO.

\vspace{-1.5em}
\begin{appendices}
\section{The Computation of Jacobian Matrix for Measurement Model} 
\label{appendix:jacobian_of_observation}

For simplicity, in the following derivations, we omit the indices for CPIs, i.e., $t$, and the parameters in brackets.
The Jacobian matrix can be expressed as
\begin{align}
    \mathbf{J}=\frac{\partial h\left( \boldsymbol{\xi } \right)}{\partial \boldsymbol{\xi }}=\left[ \frac{\partial h\left( \boldsymbol{\xi } \right)}{\partial x_{\mathrm{w}}},\frac{\partial h\left( \boldsymbol{\xi } \right)}{\partial y_{\mathrm{w}}},\frac{\partial h\left( \boldsymbol{\xi } \right)}{\partial v_x},\frac{\partial h\left( \boldsymbol{\xi } \right)}{\partial v_y} \right] ^{\mathrm{T}}. \tag{A-1} \label{eq:A-1}
\end{align}
For the partial derivatives with respect to positions, letting $\alpha_1 = x_{\mathrm{w}}$ and $\alpha_2 = y_{\mathrm{w}}$, $\frac{\partial h\left( \boldsymbol{\xi } \right)}{\partial \alpha_i }$ for $i=\{1, 2\}$ is computed by
\begin{align}
\frac{\partial h\left( \boldsymbol{\xi } \right)}{\partial \alpha _i}=\frac{\partial \mathbf{V}^{\mathrm{T}}\mathbf{a}_{\mathrm{r}}\mathbf{a}_{\mathrm{t}}^{\mathrm{T}}\mathbf{Gc}}{\partial \alpha _i}=\mathbf{V}^{\mathrm{T}}\frac{\partial \mathbf{a}_{\mathrm{r}}\mathbf{a}_{\mathrm{t}}^{\mathrm{T}}}{\partial \alpha _i}\mathbf{Gc}
, \tag{A-2} \label{eq:A-2}
\end{align}
Then, $\frac{\partial \mathbf{a}_{\mathrm{r}}\mathbf{a}_{\mathrm{t}}^{\mathrm{T}}}{\partial \alpha _i}$ can be expressed as
\begin{align}
    \frac{\partial \mathbf{a}_{\mathrm{r}}\mathbf{a}_{\mathrm{t}}^{\mathrm{T}}}{\partial \alpha _i}=\frac{\partial \mathbf{a}_{\mathrm{r}}}{\partial \alpha _i}\mathbf{a}_{\mathrm{t}}^{\mathrm{T}}+\mathbf{a}_{\mathrm{r}}\frac{\partial \mathbf{a}_{\mathrm{t}}^{\mathrm{T}}}{\partial \alpha _i}, \notag
\end{align}
For simplicity, we utilize $\mathbf{a}$ to uniformly represent $\mathbf{a}_{\mathrm{r}}$ and $\mathbf{a}_{\mathrm{t}}$ in what follows.
Additionally, we decouple $\mathbf{a}$ into a real-valued pathloss vector $\mathbf{u}$ and a complex-valued phase-shift vector $\mathbf{a}^\prime$, i.e., $\mathbf{a} = \mathbf{u} \odot \mathbf{a}^\prime$.
Therefore, we have 
\begin{align}
    \frac{\partial \mathbf{a}}{\partial \alpha _i}&=\frac{\partial}{\partial \alpha _i}\mathbf{u}\odot \mathbf{a}^{\prime}\odot \mathbf{b} \notag \\
    &=\frac{\partial \mathbf{u}}{\partial \alpha _i}\odot \mathbf{a}^{\prime}\odot \mathbf{b}+\mathbf{u}\odot \frac{\partial \mathbf{a}\prime}{\partial \alpha _i}\odot \mathbf{b}+\mathbf{u}\odot \mathbf{a}^{\prime}\odot \frac{\partial \mathbf{b}}{\partial \alpha _i}, \notag
\end{align}
whose terms can be expressed as
\begin{align}
    \frac{\partial \mathbf{u}}{\partial \alpha _i}&=-\left[ \frac{\alpha _i-\left[ \mathbf{p}_1 \right] _i}{\left\| \mathbf{d} \right\| _{2}^{3}},...,\frac{\alpha _i-\left[ \mathbf{p}_{MN} \right] _i}{\left\| \mathbf{d} \right\| _{2}^{3}} \right] ^{\mathrm{T}}, \notag \\
    \frac{\partial \mathbf{a}^\prime}{\partial \alpha _i}&=-\mathrm{j}k_0\tilde{\mathbf{a}}\odot \frac{\partial \mathbf{d}_{}}{\partial \alpha _i}, ~~\frac{\partial \mathbf{b}}{\partial \alpha _i}=-\mathrm{j}k_0\Delta T\mathbf{b}\odot \frac{\partial \hat{\mathbf{d}}}{\partial \alpha _i}, \notag
\end{align}
where $M=M_{\mathrm{t}}$ and $N=N_{\mathrm{t}}$ holds for the transmitter side and $M=M_{\mathrm{r}}$ and $N=N_{\mathrm{r}}$ holds for the receiver side, respectively.
In the above equation, the partial derivatives are given by
\begin{align}
    \frac{\partial \mathbf{d}_{}}{\partial \alpha _i}&=\left[ \frac{\partial \left\| \boldsymbol{\alpha }-\mathbf{p}_1 \right\| _2}{\partial \alpha _i},...,\frac{\partial \left\| \boldsymbol{\alpha }-\mathbf{p}_{MN} \right\| _2}{\partial \alpha _i} \right] , \notag \\
    \frac{\partial \hat{\mathbf{d}}}{\partial \alpha _i}&=\left[ \frac{\partial \mathbf{v}^{\mathrm{T}}\hat{\mathbf{d}}_1}{\partial \alpha _i},...,\frac{\partial \mathbf{v}^{\mathrm{T}}\hat{\mathbf{d}}_{MN}}{\partial \alpha _i} \right] , \notag 
\end{align}
where $\boldsymbol{\alpha}\triangleq[\alpha_1, \alpha_2]$ and
\begin{align}
    \frac{\partial \left\| \boldsymbol{\alpha }-\mathbf{p}_n \right\| _2}{\partial \alpha _i}&=\frac{\alpha _i-\left[ \mathbf{p}_n \right] _i}{\left\| \boldsymbol{\alpha }-\mathbf{p}_n \right\| _2} \notag \\
    \frac{\partial \mathbf{v}^{\mathrm{T}}\hat{\mathbf{d}}_n}{\partial \alpha _i}&=\frac{\partial}{\partial \alpha _i}\frac{\mathbf{v}^{\mathrm{T}}\left( \boldsymbol{\alpha }-\mathbf{p}_n \right)}{\left\| \boldsymbol{\alpha }-\mathbf{p}_n \right\| _2}\notag \\
    &=\frac{v_i\left\| \boldsymbol{\alpha }-\mathbf{p}_n \right\| _{2}^{2}-\left( \alpha _i-\left[ \mathbf{p}_n \right] _i \right) \mathbf{v}^{\mathrm{T}}\mathbf{d}_n}{\left\| \boldsymbol{\alpha }-\mathbf{p}_n \right\| _{2}^{3}}. \notag
\end{align}
For the partial derivatives with respect to velocities, we denote $\beta_1 = v_x$ and $\beta_2 = v_y$.
Therefore, for $i\in\{1, 2\}$, we have 
\begin{align}
    \frac{\partial h\left( \boldsymbol{\xi } \right)}{\partial \beta _i}&=\frac{\partial \mathbf{V}^{\mathrm{T}}\mathbf{a}_{\mathrm{r}}\mathbf{a}_{\mathrm{t}}^{\mathrm{T}}\mathbf{Gc}}{\partial \beta _i} =\mathbf{V}^{\mathrm{T}}\frac{\partial \mathbf{a}_{\mathrm{r}}\mathbf{a}_{\mathrm{t}}^{\mathrm{T}}}{\partial \beta _i}\mathbf{Gc}. \notag
\end{align}
Then, partial derivatives $\frac{\partial \mathbf{a}_{\mathrm{r}}\mathbf{a}_{\mathrm{t}}^{\mathrm{T}}}{\partial \beta _i}$ are given by
\begin{align}
    \frac{\partial \mathbf{a}_{\mathrm{r}}\mathbf{a}_{\mathrm{t}}^{\mathrm{T}}}{\partial \beta _i}=\frac{\partial \mathbf{a}_{\mathrm{r}}}{\partial \beta _i}\mathbf{a}_{\mathrm{t}}^{\mathrm{T}}+\mathbf{a}_{\mathrm{r}}\frac{\partial \mathbf{a}_{\mathrm{t}}^{\mathrm{T}}}{\partial \beta _i}. \notag 
\end{align}
Again, we utilize $\mathbf{a}$ as a unified notation.
Given $\boldsymbol{\beta}\triangleq[\beta_1, \beta_2]$, we have
\begin{align}
    \frac{\partial \mathbf{a}}{\partial \beta _i}&=\frac{\partial \left( \tilde{\mathbf{a}}\odot \mathbf{b} \right)}{\partial \beta _i}=\tilde{\mathbf{a}}\odot \frac{\partial \mathbf{b}}{\partial \beta _i}, \notag \\
    \frac{\partial \mathbf{b}}{\partial \beta _i}&=-\mathrm{j}k_0\Delta T\mathbf{b}\odot \left[ \frac{\beta _i-\left[ \mathbf{p}_1 \right] _i}{\left\| \boldsymbol{\beta }-\mathbf{p}_1 \right\| _2},...,\frac{\beta _i-\left[ \mathbf{p}_{MN} \right] _i}{\left\| \boldsymbol{\beta }-\mathbf{p}_{MN} \right\| _2} \right]. \notag
\end{align}
Here, the derivation of the Jacobian matrix of the observation model is complete.

\vspace{-1.5em}
\section{The proof of \textbf{Lemma \ref{lemma:1}}} \label{appendix:subspace}
Before proceeding with this proof, we first define the parallel subspace as $V^{\parallel}$ and the orthogonal subspace as $V^{\bot}$.
In this proof, we resort to the proof by contradiction.
In particular, we assume that part of the optimal $\mathbf{q}^\star$ can be located in the orthogonal subspace $V^{\parallel}$.
Then, we prove that we can always construct a better solution that is not only included in the parallel subspace $V^{\parallel}$ but also achieves a higher objective value.
This solution contradicts the initial assumption of optimality, thereby leading to the conclusion $\tilde{\mathbf{q}}^{\star} \in V^{\parallel}$.

Following this logical flow, we first assume that the optimal solution $\tilde{\mathbf{q}}^{\star} = \tilde{\mathbf{q}}_\perp + \tilde{\mathbf{q}}_\parallel$, where $\tilde{\mathbf{q}}_\perp \in V^{\perp}$ and $\tilde{\mathbf{q}}_\parallel \in V^{\parallel}$.
Then, the objective function is given by 
\begin{align}
    c\left( \mathbf{X} \right)\frac{I_1}{I_2}=c\left( \mathbf{X} \right) \frac{( \tilde{\mathbf{q}}_{\bot}+\tilde{\mathbf{q}}_{\parallel} ) ^{\mathrm{H}}\mathbf{A}\left( \mathbf{X} \right) ( \tilde{\mathbf{q}}_{\bot}+\tilde{\mathbf{q}}_{\parallel} )}{( \tilde{\mathbf{q}}_{\bot}+\tilde{\mathbf{q}}_{\parallel} ) ^{\mathrm{H}}\mathbf{B}\left( \mathbf{X} \right) ( \tilde{\mathbf{q}}_{\bot}+\tilde{\mathbf{q}}_{\parallel})}. \tag{B-1} \label{eq:B-1}
\end{align}
Note that as $\tilde{\mathbf{q}}_\perp \in V^{\perp}$, we have $\tilde{\mathbf{q}}_\perp \perp \mathbf{H}_{\mathrm{w}}^{} (\hat{\mathbf{v}}_{\mathrm{w}},\hat{\mathbf{r}}_{\mathrm{w}})$ and $\tilde{\mathbf{q}}_\perp \perp \mathbf{h}_{\mathrm{b}}^{}\left( \mathbf{r}_{\mathrm{b}},\mathbf{X} \right)$.
The numerator of \eqref{eq:B-1} is computed by
\begin{align}
    I_1 = I_{11} + I_{12} + I_{13} + I_{14}, \tag{B-2} \label{eq:B-2}
\end{align}
where the terms are expressed as 
\begin{align}
    I_{11}&=\tilde{\mathbf{q}}_{\bot}^{\mathrm{H}}\mathbf{A}\left( \mathbf{X} \right) \tilde{\mathbf{q}}_{\bot}^{}=-\Gamma _{\mathrm{sen}}\| \tilde{\mathbf{q}}_{\bot}^{} \| _{2}^{2}, \notag \\
    I_{12}&=\tilde{\mathbf{q}}_{\bot}^{\mathrm{H}}\mathbf{A}\left( \mathbf{X} \right) \tilde{\mathbf{q}}_{\parallel}^{}=0, \notag \\
    I_{13}&=\tilde{\mathbf{q}}_{\parallel}^{\mathrm{H}}\mathbf{A}\left( \mathbf{X} \right) \tilde{\mathbf{q}}_{\bot}^{}=0, ~~I_{14}=\tilde{\mathbf{q}}_{\parallel}^{\mathrm{H}}\mathbf{A}\left( \mathbf{X} \right) \tilde{\mathbf{q}}_{\parallel} > 0. \notag
\end{align}
Similarly, the denominator of \eqref{eq:B-1} can be simplified by 
\begin{align}
    I_2=\tilde{\mathbf{q}}_{\parallel}^{\mathrm{H}}\mathbf{B}\left( \mathbf{X} \right) \tilde{\mathbf{q}}_{\parallel}^{}. \tag{B-3} \label{eq:B-3}
\end{align}
Therefore, in light of \eqref{eq:B-2} and \eqref{eq:B-3}, the objective function can be written as
\begin{align}
    \eqref{eq:B-1} =\frac{\tilde{\mathbf{q}}_{\parallel}^{\mathrm{H}}\mathbf{A}\left( \mathbf{X} \right) \tilde{\mathbf{q}}_{\parallel}^{}-\Gamma _{\mathrm{sen}}\left\| \tilde{\mathbf{q}}_{\bot}^{} \right\| _{2}^{2}}{\tilde{\mathbf{q}}_{\parallel}^{\mathrm{H}}\mathbf{B}\left( \mathbf{X} \right) \tilde{\mathbf{q}}_{\parallel}^{}}. \tag{B-4} \label{eq:B-4}
\end{align}
From \eqref{eq:B-4}, we can see that the perpendicular component $\tilde{\mathbf{q}}_{\perp}$ actually diminishes the value of the objective function.
Keeping this in mind, a better solution can be constructed by allocating all power into the parallel component, which yields 
\begin{align}
    \tilde{\mathbf{q}}^{\star \star}=\sqrt{\frac{\left\| \tilde{\mathbf{q}}^{\star} \right\| _{2}^{2}}{\left\| \tilde{\mathbf{q}}^{\star} \right\| _{2}^{2}-\left\| \tilde{\mathbf{q}}_{\bot}^{} \right\| _{2}^{2}}}\tilde{\mathbf{q}}_{\parallel}^{}. \tag{B-5} \label{eq:B-5}
\end{align}
By adopting this new solution, the new objective value can reach $\frac{\tilde{\mathbf{q}}_{\parallel}^{\mathrm{H}}\mathbf{A}\left( \mathbf{X} \right) \tilde{\mathbf{q}}_{\parallel}^{}}{\tilde{\mathbf{q}}_{\parallel}^{\mathrm{H}}\mathbf{B}\left( \mathbf{X} \right) \tilde{\mathbf{q}}_{\parallel}^{}} > \eqref{eq:B-1} $, which indicates our initial assumption on optimal solution is contradicted.
Thus, the optimal $\tilde{\mathbf{q}}^{\star}$ is entirely included in the parallel subspace $V^{\parallel}$.
Here, this proof is completed.
\end{appendices}
\bibliographystyle{ieeetr}
\begin{spacing}{.96}
    \bibliography{reference.bib}
\end{spacing}
\end{document}